\documentclass[amssymb,amsfonts,aps,prb,twocolumn,longbibliography]{revtex4-2}

\usepackage{graphicx}
\usepackage{dcolumn}
\usepackage{bm}
\usepackage{bbold}
\usepackage[normalem]{ulem}
\usepackage{xcolor}
\usepackage{comment}
\usepackage{multirow}
\usepackage{amsmath}
\usepackage{siunitx}
\usepackage{tabularx}
\usepackage{dsfont}
\usepackage{float}
\usepackage{braket}
\usepackage{placeins}
\usepackage[shortlabels]{enumitem}

\usepackage{appendix}

\allowdisplaybreaks

\DeclareSIUnit\angstrom{\text{Å}}

\begin{document}

\title{Landau quantization of $C_3$ high-fold fermions in honeycomb lattices}

\author{
L. Madail$^{1,2}$, J. Fern\'{a}ndez-Rossier$^{1}$\thanks{On permanent leave from Departamento de F\'{i}sica, Universidad de Alicante, 03690 San Vicente del Raspeig, Spain}\thanks{joaquin.fernandez-rossier@inl.int}
}
\affiliation{
\\
$^1$ International Iberian Nanotechnology Laboratory (INL), Av. Mestre Jos\'{e} Veiga, 4715-330 Braga, Portugal
\\
$^2$ Department of Physics and I3N, University of Aveiro, 3810-193 Aveiro, Portugal}

\date{\today}

%%%%%%%%%%%%%%%%%%%%%%%%%%%%%%%%%%%%%%%%%%%%%%%%%%%%%%%%%%

\begin{abstract}

We present the theory of  the Landau levels of $C_3$ high-fold fermions
in  honeycomb lattices. These fermions arise as a generalization of the tight-binding model in a honeycomb lattice with one orbital per site,  known to describe graphene, to a family of models with
several orbitals per site transforming according to representations of $C_3$, relevant for triangulene two-dimensional crystals.  
The models are defined by the number of orbitals $n_a$ and $n_b$ at the $a$ and $b$ sites of the unit cell. They feature energy bands that combine flat bands and Dirac cones, that are isomorphic to those of graphene, the Lieb, $p_{xy}$ honeycomb,  and the Kagome tight-binding models. Combining tight-binding and $k\cdot p$ methods, here we compute the Landau levels of the $(1,1)$, $(1,2)$, $(2,2)$ and $(3,3)$ models, we compare them to those of the reference models, and we explore the similarities and differences with the Landau levels in graphene. Our findings reveal that the nature of $C_3$ high-fold fermions combine Dirac-type Landau quantization, originating from the linear crossings and associated relativistic band dispersion, with anomalous Landau quantization rooted in the quantum geometry of the singular flat bands.

\end{abstract}

\maketitle

\section{Introduction}
Many of the outstanding electronic properties of graphene can be associated to the presence of Dirac cones in its energy bands and to  peculiar properties of their Bloch states \cite{Neto2009}. Non-generic electronic properties of graphene include an electron-hole symmetric Landau level (LL) spectrum that scales with $\sqrt{n+1}$, where $n=0,1,2,...$ plus a $E=0$ Landau level, pseudo-diffusive ballistic transport \cite{Tworzydlo2006}, Klein tunneling \cite{Katsnelson2006,Beenakker2008}, peculiar $E=0$ states associated to point-defects \cite{Pereira2006} and zigzag edge states \cite{Brey2006}. 
All of these properties are accounted for by a tight-binding (TB) model with a single orbital in a honeycomb lattice.

In  recent papers \cite{Ortiz2022,Madail2024} we introduced a generalization of the graphene tight-binding model to a class of cases where every site can host several orbitals, with the condition that they have $C_3$ symmetry. Specifically, we considered lattices where the two sites of the unit cell host $n_a, n_b$ orbitals with $C_3$ symmetry, where $n_a$ and $n_b$ are integers larger than zero. We refer to these as high-fold $C_3$ fermions. Thus, the case of graphene is a particular instance of this generalization, with  $n_a=n_b=1$.  The instances with $n_{a,b}>1$ arise naturally in the description of two-dimensional crystals of graphene triangulenes \cite{Ortiz2022,Catarina2023} as well as networks of topoplogical graphene nanoribbon junctions \cite{Tamaki2020} and give rise to a variety of very interesting energy bands.
Throughout this work, parentheses $(n_a,n_b)$ specify the number of zero modes associated with the $a$ and $b$ sites of the honeycomb lattice, whereas brackets $[n_a,n_b]$ specify the sizes of the corresponding triangulenes, measured by the number of benzenes along each edge. Since an $[n]$triangulene hosts $(n-1)$ zero modes, the $(n_a,n_b)$ model describes the low-energy sector of the $[n_a+1,n_b+1]$triangulene crystal.

The (1,2) model realizes a spin-1 Dirac crossing with the same low-energy Hamiltonian and Landau quantization as the Lieb lattice \cite{Weeks2010,Tsai2015,Jakubsky2022}, with a $E=0$  flat band degenerate at the $\Gamma $ point with a single Dirac cone. The bands of $(2,2)$ are isomorphic to the $p_x-p_y$ honeycomb model \cite{Wu2007}, where
two electron-hole symmetric flat bands encapsulate 2 graphene-like bands.  Unlike the others, the $(n_a=3,n_b=3)$ $C_3$ high-fold fermion in the honeycomb lattice has two sets of Kagome-like  bands separated by a gap \cite{Delgado2023, Sethi2021, Fang2024}.  Both the $(2,2)$ and $(3,3)$ feature the so called singular flat bands, where a parabolic band touches a flat band at the $\Gamma$ point \cite{Rhim2020,Rhim2021}.
Physical realizations of this type of crystals include not only triangulene 2D crystals \cite{Ortiz2022,Catarina2022,Madail2024}, but also artificial lattices made with adatoms \cite{Sierda2023} and molecules \cite{Gomes2012} on surfaces, photonic crystals \cite{Plotnik2013}, polaritonic crystals \cite{Jacqmin2014}, graphene nanoribbon junctions \cite{Tamaki2020} and electric circuits \cite{Chen2023}.  

In this work we study the electronic structure of high-fold $C_3$ fermions under the influence of a strong
perpendicular magnetic field, focusing on the emergence of Landau levels and their properties.  The differences between Landau levels of electrons in graphene with those of Schrödinger electrons in a two-dimensional electron gas \cite{Novoselov2005,Zhang2005} settled the notion that low-energy electrons in graphene are governed by a Dirac equation.   An obvious question is wether the LL associated with the Dirac cones present in the $C_3$ high-fold fermions will be similar to those of graphene. In addition, Landau levels associated to singular flat bands are also very different from conventional LL. In the case of the $(1,2)$ system, that features a signle electron-hole symmetric  pair Dirac cones at the $\Gamma$ point, intersected by a $E=0$ flat band,  it is interesting to see whether the $E=0$ Landau level expected for Dirac fermions is coupled to the pre-existing $E=0$ flat band.

The rest of this paper is organized as follows. In section \ref{methods} we present the three different methods used to compute the Landau level structure.  These include a full tight-binding model for a ribbon that yields both the bulk LL as well as the chiral edge states, an effective reduced tight-binding model capturing the low-energy physics and  $k\cdot p$ theory, that yields the bulk LL. Sections \ref{22}, \ref{23}, \ref{33} and \ref{44} present the magnetic quantization signatures for the $(1,1)$, $(1,2)$, $(2,2)$ and $(3,3)$. 
In Section \ref{sec_final}, we summarize our main conclusions, highlighting how $C_3$ high-fold fermions in multiorbital honeycomb lattices combine Dirac-type with anomalous Landau quantization behavior.

\section{Methods\label{methods}}

We use three different methods to compute the LL of the  $C_3$ high-fold fermions in honeycomb lattices. First, the full tight-binding Hamiltonian that describes a triangulene two-dimensional crystal \cite{Ortiz2022,Madail2024}, where the basis set is given by the $\pi$ orbitals of the carbon sites of the triangulenes.  Second, a reduced TB model, where the basis set is given by the zero-modes of the triangulenes. 
Triangulenes with $n_a$ benzenes per side, denoted as 
[$n_a$]triangulene host $n_a-1$ half-filled zero modes \cite{FernandezRossier2007,Ortiz2019}.  
In both tight-binding models, the
effect of an out-of-plane magnetic field is included through a Peierls complex phase factor that captures the influence of the magnetic vector potential. As a result of the Peierls susbstituion, the Hamiltonian loses translational invariance along the $y$ axis. Therefore, we compute the electronic structure of ribbons, infinite along the $x$ axis.
The crystal momentum $k_x$ is expressed in units of $2\pi/a$, where $a$
is the lattice period of the $[n_a,n_b]$triangulene crystal along the periodic direction, and  can be expressed in terms of the  carbon-carbon distance $a_{CC}$ as
$a\equiv(n_a+n_b+1)\sqrt{3}a_{CC}$. 
This approach gives a set of one-dimensional energy bands consisting of flat bulk Landau level bands, arising from states localized away from the ribbon edges, together with dispersive edge states.

The third method is a $k\,\cdot\,p$ theory \cite{Luttinger1951,Katsnelson2006a} obtained by Taylor expansion of the TB Hamiltonian around either the $\Gamma$ or the $K,K'$ points of the Brillouin zone of the honeycomb lattice.  The effect of the magnetic field is obtained through the Peierls substitution  $\vec{\Pi}\equiv\vec{p}-e\vec{A}$.

%The relation between triangulene 2D crystals and the $C_3$ high-fold fermions is the following: 
%Triangulene 2D crystals are honeycomb lattices with a unit cell with a $[n_a]$triangulene in the A site and a $[n_b]$triangulene in the $B$ site. Since each monomer has $n_{a/b-1}$ zero modes, the $(n_a+n_b-2)$ bands around zero-energy constitute the generalization of the graphene model to the case of $C_3$ high-fold fermions \cite{Madail2024} with ($n_a-1,n_b-1$) states in the $A,B$ triangular sublattices. Just as in our previous work \cite{Madail2024}, we use brackets to refer to the number of benzenes on a triangulene edge  and parenthesis to count the number of states in a given site of the honeycomb lattice.

\subsection{Full tight-binding}
\begin{figure}
    \centering
    \includegraphics[width=1\linewidth]{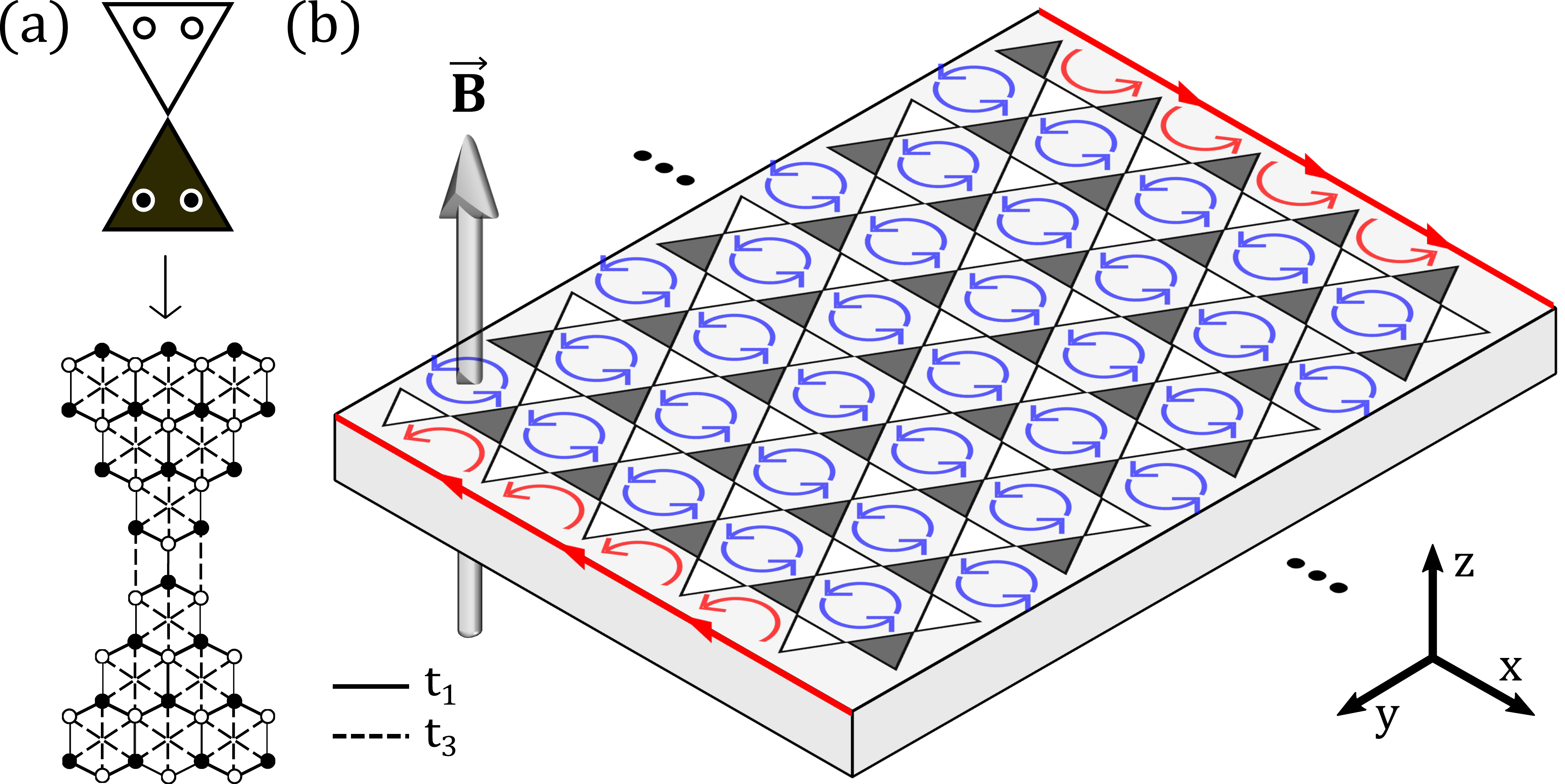}
    \caption{Schematics of a triangulene ribbon under the influence of an out-of-plane magnetic field. (a) Example of a dimer composed of two orbitals per site accomplished by coupling two $[3]$triangulenes tip-to-tip. (b) Ribbon configuration with translation invariance in the $x$-direction, zigzag open boundaries in the $y$-direction and out-of-plane magnetic field. }
    \label{fig:1}
\end{figure}

The single particle properties of $[n_a,n_b]$triangulenes are described by the following bipartite tight-binding Hamiltonian \cite{Ortiz2022}   
\begin{equation}
    H_{\text{TB}}=-t_1\sum_{<i,j>}a^\dagger_i b_j-t_3\sum_{\lll i,j\ggg}a^\dagger_i b_j,
    \label{eq:1}
\end{equation}
where $a^\dagger_i$ ($b^\dagger_i$) is the fermionic operator that creates an electron at site $i$ belonging to the $A$($B$)-sublattice (see Fig.~\ref{fig:1}a). The unit cell dimer is composed by two triangulenes of A and B types with a single $p_z$ orbital per carbon atom. The first (second) sum is taken between first (third) neighbor sites in each triangulene and includes all intercell hopping elements at the binding sites of the dimer. 
Following previous work \cite{Ortiz2022}, we assume $t_2=0$, so that the model is bipartite ensuring thereby local charge neutrality, in agreement with density functional based calculations \cite{Ortiz2022,Catarina2023,Madail2024}. 

The single-particle spectrum of individual $[n]$triangulenes  described with Eq.~\eqref{eq:1} has a set of electron-hole symmetric states with finite-energy, well separated by a gap, and $(n-1)$ in-gap sublattice-polarized zero modes  with $E=0$. %These modes can be chosen to be eigenvectors of the $C_3$ symmetry operator, on account of the point symmetry of triangulenes. 
Intermolecular hybridization among the zero modes of adjacent triangulenes can only be established by third-neighbor hopping, the leading term that produces intermolecular hybridization. 
It was found \cite{Ortiz2022} that a good agreement between spin-unpolarized density functional theory (DFT) and tight-binding energy bands of $[n_a,n_b]$triangulene 2D crystals is obtained if we take $t_3=0.1t_1$. These calculations display a wealth of weakly dispersive half-filled narrow bands located around the Fermi energy, well separated from higher/lower energy bands. Specifically, we emulate the graphene spectrum \cite{Tran2017} for $n_a = n_b = 2$, spin-1 Dirac electrons \cite{Mizoguchi2021} and the Lieb lattice \cite{Weeks2010,Tsai2015} for $n_a = 2,n_b = 3$, $p_{x,y}$-orbital honeycomb physics \cite{Wu2007, Congjun2008} for $n_a = n_b = 3$, and two kagome-like bands \cite{Fang2024, Delgado2023} separated by a gap with a flat valence and conduction bands for $n_a = n_b = 4$.

Under the influence of a perpendicular magnetic field $\vec{B}=B\hat{z}$, we expect the formation of Landau levels and edge states (see Fig.~\ref{fig:1}b). The effect of the external field is included using the Peierls substitution
\begin{equation}
    t_{i,j} \rightarrow t_{i,j}e^{i\phi_{i,j}},
\end{equation} 
where the phase $\phi_{i,j}$ that modulates the hopping term \cite{Peierls1933, Thouless1982} reads
\begin{equation} 
\phi_{i,j}=\frac{e}{\hbar} \int_i^j\vec{A}.d\vec{r}\,.
\end{equation}
This approach \cite{Wakabayashi2001,Katsnelson2006, Xiao2007,Lado2013} provides a complete description of the non-interacting electrons under $\vec{B}$. The gauge choice $\vec{A}=-By\hat{x}$ allows one to preserve the original periodicity of the crystal in the $x$-direction and study one-dimensional stripes that can fully capture bulk along with edge solutions. Then, the phase dependence on the magnetic field follows 
\begin{equation}
    \phi_{i,j}=-\frac{eB}{2\hbar}(x_j-x_i)(y_j+y_i)
\end{equation}
where $(x_i,y_i)$ is the real-space position of atom $i$. The magnetic field introduces a new length scale $\ell_B=\sqrt{\frac{\hbar}{eB}}$. To ensure both localization of the bulk states and dispersive states occurring only at the boundary, we assume a sufficiently high magnetic field so that $l_{B}<W$ where W is the width of the ribbon.  

The Hamiltonian of the zigzag $[n_a,n_b]$triangulene ribbon with $N_{td}$ dimers in the unit cell can be expressed by the banded block-Toeplitz matrix described in \cite{Madail2024}. The number of energy bands captured by numerical diagonalization of this TB Hamiltonian under the influence of an external magnetic field with a unit cell composed of $N_{td}$ dimers is determined by the total number of atoms $N_t=\left[(n_a+2)^2+(n_b+2)^2-6\right]N_{td}$, as the number of carbon sites of an $[n]$triangulene is $(n+2)^2-3$. In the low-energy window, $(n_a+n_b-2)N_{td}$ bands $\epsilon_n(k_x)$ around $E=0$ were found to be embedded inside a large gap.

\subsection{Reduced tight-binding}
Since we are interested in the low-energy bands, whose Bloch states are almost entirely made of the zero modes of single triangulenes \cite{Ortiz2022,Madail2024}, we can employ a reduced tight-binding model expressed in the basis of zero modes.  This has the advantage of reducing dramatically the size of the Hilbert space. The quality of the approximation can be benchmarked by comparing with the complete tight-binding described in the previous section. In the zero-mode basis the effective 2D Bloch Hamiltonian can be expressed in the basis of the $n_a+n_b-2$ zero modes as \cite{Madail2024}
\begin{equation}
	H_{\text{eff}}(\vec{k})=
	\begin{pmatrix}
		0_{n_a-1} & \tau(\vec{k})\\
		\tau^\dagger(\vec{k})& 0_{n_b-1}
	\end{pmatrix}
 \label{H2Def}
\end{equation}
describing the coupling between $(n_a-1)\ket{\psi_{a}}$ and $(n_b-1)\ket{\psi_{b}}$ modes that populate sites of individual triangulenes belonging to the $A$ and $B$ sublattice, respectively (see Fig.~\ref{fig:1}a). 
The elements of the $(n_a-1)\times(n_b-1)$ off-diagonal matrix $\tau(\vec{k})$ are given by
\begin{equation}
    \tau_{ab}(\vec{k})=\sum_{m=0,1,2}\tau^{ab}_{m}, \qquad \tau^{ab}_{m}=e^{i\vec{k}.\vec{R}_m}t_{m}^{ab},
    \label{eq:3}
\end{equation}
determined using the null vector associated to the intracell hopping terms $R_0  = (0,0)$ and the primitive vectors $\vec{R}_{1,2}=(n_a+n_b+1) \vec{a}_{1,2}$ that define the hexagonal lattice of the triangulene dimer in terms of the graphene primitive vectors $\vec{a}_{1,2}$. All matrix elements can be determined by the components of the zero mode wave functions at the binding sites
\begin{equation}
    t^{ab}_{m}=t_3\sum_{i,j\big|_m}\psi_a(i)\psi_b^*(j))
    \label{tabm}
\end{equation}
where $\psi_{a}(i)=\braket{i|\psi_{a}}$ are the components of the $A$-triangulene zero mode wave function at the atomic site $i$. The sum in Eq.~(\ref{tabm}) runs between adjacent triangulenes at corner $m$ over the pair of binding sites $i$ and $j$ on the A and B monomer, respectively. Moreover, intercell ($m=1,2$) contributions can be expressed in terms of the intracell ($m=0$) contributions by $
t_{1,2}^{ab}=e^{\pm i(\omega_b-\omega_a)}t_{0}^{ab}$ where $\omega_{a/b}$ are the phases of the zero modes upon a $R_{2\pi/3}$ rotation operation. This stems from the rotational degree of freedom between states in the same sublattice associated with $C_3$ symmetry.

The non-zero contributions of the zigzag $[n_a,n_b]$triangulene ribbon Hamiltonian in this reduced version can be retrieved from the block matrices of Eq.~(\ref{H2Def}). Following \cite{Madail2024}, intra-dimer contributions will be exclusively intra-cell and independent on $k$ and inter-dimer contributions contain both intra-cell and inter-cell terms of the parent Hamiltonian.

In order to correctly capture the influence of the magnetic field on the ribbon we incorporate the effect of $\vec{B}$ on the calculation of the new effective hopping terms as \cite{Ortiz2022}

\begin{equation}
\tilde{t\,}^{ab}_{m,n_y}=\sum_{i,j\big|_{m,n_y}}\tilde{t}_3\tilde{\psi}_a(i)\tilde{\psi}_b^*(j)
\end{equation}
which will now be dependent on the vertical position index $n_y$ within the ribbon unit cell due to the choice of vector potential. We apply the Peierls substitution for the modulation of the third neighbour hopping term in 

\begin{equation}
\tilde{t}_3=t_3e^{i\phi_{ij|_{m,n_y}}}=t_3e^{i\phi_{ij|_m}}e^{i\phi_y n_y}
\end{equation}
where
$\phi_y=-\frac{eB}{\hbar}R_y \Delta x_{ij}$
 is the position dependent phase correction, $\Delta x_{ij}=(x_j-x_i)$ and $R_y=||\vec{R_1}+\vec{R_2}||$.

%Note that the magnetic field will also influence the weight of the zero modes $\ket{\tilde{\psi}_{a/b}}$ of individual triangulenes and this will also depend on the vertical position of such triangulene in the ribbon. 
The magnetic field modifies the molecular zero modes, while their representation in the chosen Landau gauge also depends on the monomer vertical position.
Thus, we determine the zero mode solutions of a triangulene monomer TB Hamiltonian at an arbitrary position in real space with the inclusion of a Peierls substitution in the leading (first neighbor) hopping term. We can do this in two different manners.  When possible (see Appendix~\ref{SMA}), we take advantage of the sublattice imbalance $(n_{a/b}-1)$ of $[n_{a/b}]$triangulenes to find the analytical magnetic field dependent solutions. These can be retrieved by solving the eigenvalue equation $H_{TB}^{B\neq0}\ket{\tilde{\psi}_{a/b}}=0$ at sites in the minority sublattice, where the weight of the zero modes vanish. One can also numerically find the zero mode solutions by diagonalizing the Hamiltonian for each monomer positioned in the ribbon. For convenience, we recover solutions of the form $\ket{\tilde{\psi}_{a,b}}=\ket{\psi_{a,b}(n_y)}$ as a function of the position index $n_y$ of the monomer in the unit cell.

\subsection{$k\,\cdot\,p\,$ theory} \label{SecII.C}

The Landau levels of triangulenes can be calculated using the $k\,\cdot\,p$ theory \cite{Luttinger1951,Katsnelson2006a}. To do so, we proceed in two steps. First, 
we consider a high-symmetry point $\mathbf{k}_0$, such as   as the $K$ or the $\Gamma$ points, and write
$\mathbf{k}=\mathbf{k}_0+\mathbf{q}$. The Bloch Hamiltonian is
expanded as
\begin{equation}
H(\mathbf{k}_0+\mathbf{q})
=
H^{(0)}+H^{(1)}(\mathbf{q})
+H^{(2)}(\mathbf{q})+\cdots ,
\label{eq:kp}
\end{equation}
where $H^{(m)}$ contains terms of order $q^m$.
The $k\cdot p$ Hamiltonian is thus a small-order polinomial in $q_x,q_y$.
Second,  we replace  $(q_x,q_y)$ by   $\frac{1}{\hbar}(\Pi_x,\Pi_y)$,  where $\vec{\Pi}\equiv\vec{p}-e\vec{A}$, so that the effective Hamiltonian becomes now a differential operator.
%It treats the behavior of electrons bound to a lattice as a perturbation in an uniform effective potential with each energy band linked to an effective mass.
This description is only valid if the degree of localization of the electronic wave functions measured by the magnetic length $\ell_B\gg \sqrt{3} \sqrt{n+1} a_{CC}$, where $n$ is the Landau level index, introduced below.
%The effect of the magnetic field is introduced through the canonical momentum operator $\vec{\Pi}\equiv\vec{p}-e\vec{A}$.
%In the low-energy regime, this amounts to replacing $\vec{q}$ with this operator in the reduced Hamiltonian $H_{\text{eff}}(\vec{q)}$ after expansion of Eq.~(\ref{H2Def}) around the center of the Brillouin zone. 
For a constant magnetic field perpendicular to the lattice plane, the Landau gauge choice $\vec{A}=-By\,\hat{x}$  allows one to assume eigenfunctions of the form $e^{iq_x x}\psi_n(q_x,y)$ and the operator $p_x$ replaced by the quantum number $\hbar q_x$. To find the solution of the Schr\"odinger equation we may introduce the ladder operators 
\begin{equation}
    a(q_x)=\frac{1}{\sqrt{2}}(Q+iP),\quad
    a^\dagger(q_x)=\frac{1}{\sqrt{2}}(Q-iP).
    \label{eq:ladder}
\end{equation}
These are defined with respect to the internal coordinates of the Landau levels given by the dimensionless canonical operators $Q(k_x)\equiv\frac{y}{\ell_B}+q_x\ell_B$,  %\redmark{(I think it is a $+$ and not a $-$ like in your paper with Lado do you confirm?) } 
%JFR: No time to check, I trust you}
$P\equiv\frac{\ell_B}{\hbar}p_y$, and satisfy  $[a(q_x),a(q_x)^\dagger]=1$. For a generic high-symmetry point of interest $\Lambda$, we write the effective mass Hamiltonian in the low-energy limit as 
\begin{equation}
 	H_n=
    %A_0
    \varepsilon_0 \begin{pmatrix}
		0_{n_a-1} & \tilde{\tau}(a,a^\dagger)\\
		\tilde{\tau}^\dagger(a,a^\dagger)& 0_{n_b-1}
        \label{Hn}
	\end{pmatrix}
\end{equation} 	
where we have defined $\varepsilon_0=\frac{\hbar v_F}{\ell_B}$ and the $\tilde{\tau}$ matrices are first order expansions of $\tau$ around the center of the Brillouin zone after the re-definition $\vec{k}\rightarrow \vec{q}+\Lambda$.
Importantly, the  Hamiltonian of Eq.~(\ref{Hn}) does not depend explicitly on $k_x$.  This implies that its spectrum has a macroscopic degeneracy.  For a sample of dimensions $L_x, L_y$, we have $\Delta k_x= \frac{2\pi}{L_x}$ and, consequently, $\Delta y=\ell_B\Delta Q=\ell_B^2 \Delta k_x$.  The number of states that every eigenvalue of Eq.~\eqref{Hn} can accommodate is $N_{\rm states}=\frac{L_y}{\Delta y}=\frac{L_x L_y}{2\pi \ell_B^2}=\frac{A B}{\Phi_0}$, where $\Phi_0=\frac{h}{|e|}$.  This corresponds to one electron per magnetic flux quantum.

We note that inverting the sign of $B$ in the definitions above requires to redefine the operator $Q$. that leads to new definitions of ladder operators where the roles of $a$ and $a^{\dagger}$ are swapped, compared to those of Eq.~\eqref{eq:ladder}. The valley transformation exchanges the $C_3$ conjugate states and reverses the chiral momentum coupling (see Appendix~\ref{sec:valley-vs-a}). In the Landau level representation, the latter is encoded by $a\leftrightarrow a^\dagger$, thereby modifying the orbital composition of the wave functions between the two valleys.

In the case of graphene, this approach leads to a problem that can be solved analytically, as the square of the operator $H$ is diagonal \cite{peres07,Lado2013}.
In our context, the squaring operation does not diagonalize Eq.~(\ref{Hn}) and 
we resort to a numerical method whenever a simple analytical solution cannot be extracted. We expand the eigenstates of the Hamiltonian as a linear combination of vector states of the form
\begin{equation}
\ket{n,b}=\ket{n}\otimes\vec{b}_i,
\end{equation}
where $\ket{n}$ are the eigenstates of the $a^\dagger a$ operator, 
$a^\dagger a\ket{n}= n\ket{n}$, with $n=0,1,...$ and $\vec{b}$ are the unit vectors of dimension $n_a+n_b-2$, 
\begin{equation}
\begin{aligned}
\vec{b}_1= \left(\begin{array}{c} 1 \\ 0 \\...\end{array}\right),\;
\vec{b}_2 =
\left(\begin{array}{c} 0 \\ 1 \\...\end{array}\right), \;...\;,\;
 \vec{b}_{n_a+n_b-2} =  \left(\begin{array}{c} 0 \\ ... \\1\end{array}\right) 
\end{aligned} 
\end{equation}

% canonical transformation to solve the eigensystem $H_n\ket{\nu}=E_\nu\ket{\nu}$. By defining our eigenvalues in the basis $\ket{n,b}$, labeling the number states $\ket{n}$ and the standard orthonormal basis set $\ket{b_i}$ for $\mathds{R}^3$, one may write the eigenstates of the system in the form

\begin{equation}
	\ket{\nu}\equiv\sum_{n,b} C_\nu(n,b)\ket{n,b}
\end{equation} 
%C_\nu(n,i) here and C_\nu(n,b) below ? JFR: sum should be over $b$, or over $B_i$. It's a matter of notation
Ensuring both orthonormal condition $\braket{n,b\,|\,n',b'}=\delta_{nn',bb'}$ and completeness relation $\sum_{n,b}\ket{n,b}\bra{n,b}=\mathds{1}$ are satisfied, we are able to write our system as
\begin{equation}
	\sum_{n',b'}H_{bb'}(n,n')C_\nu(n',b')=E_\nu C_\nu(n,b)
	\label{Eq:Mbn}
\end{equation}
with $H_{bb'}(n,n')=\bra{n,b}H\ket{n',b'}$.
For the numerical solution of Eq.~\eqref{Eq:Mbn} we choose a finite basis set, which implies
selecting a cutoff ($N_{cf}$) and diagonalize a square matrix of dimension $(n_a+n_b-2)N_{cf}$. The choice of cutoff is taken as a compromise between two antagonistic requirements.  From the numerical point of
view, the largest the basis set the better the numerical accuracy. On the other hand, 
$\varepsilon_0 N_{cf}$ should be smaller than the bandwidth of the zero-energy bands, to comply with the hypothesis of the $k\,\cdot\,p$ approximation. 
The resulting matrix is sparse, with non-zero entries in the first upper $n'=n+j$ and lower $n'=n-j$ diagonals, with $j$ a non-negative integer no larger than the expansion order $j^{th}$. For instance, a second order expansion of the reduced Hamiltonian in Eq.~(\ref{H2Def}) yields, at most, five non-zero diagonals in $H_{bb'}(n,n')$ composed of block square matrices of order $(n_a+n_b-2)$.

\section{Landau levels in $[2,2]$triangulene \label{22}}
\begin{figure}[t]
    \centering
    \includegraphics[width=1\linewidth]{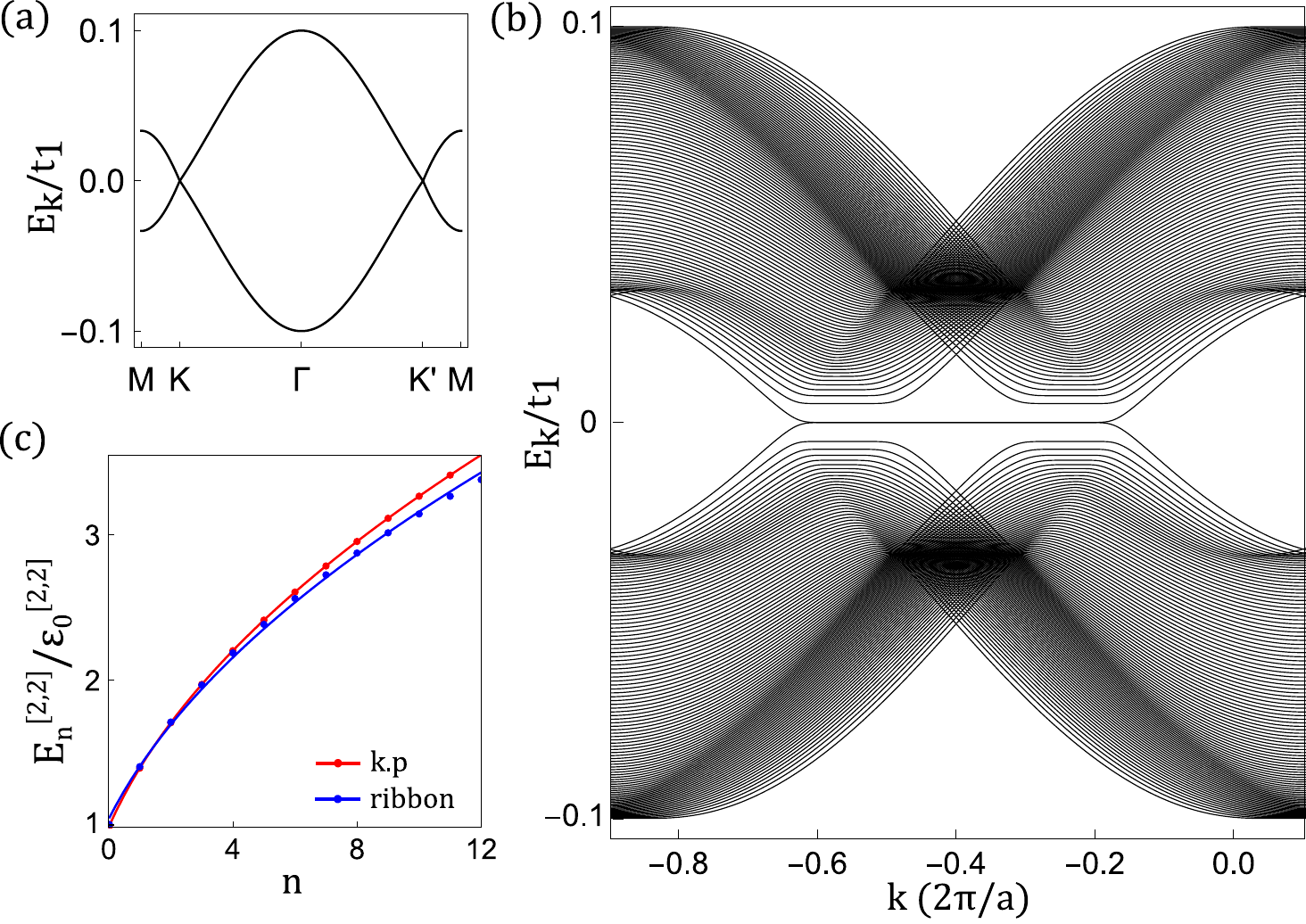}
    \caption{Landau level structure of $[2,2]$triangulenes. (a) Energy bands of the crystal obtained using the reduced TB model, with $t_3 = 0.1t_1$. (b) Energy bands of the zigzag ribbon configuration with $N_{td}=100$ dimers in the unit cell and magnetic field set to $B=6$ T. 
    We notice the presence of
    a set of Landau levels around each of the two Dirac points.
 (c) Scaling of the first Landau levels show a $\sqrt{n+1}$ behavior both in the numerics (blue) and using the effective mass approximation (red).}
    \label{fig:2}
\end{figure}

We now discuss the Landau levels of the honeycomb lattice made with  $[2,2]$triangulenes. 
These have one zero mode per triangulene, therefore, the low-energy bands are isomorphic to that of graphene \cite{Ortiz2022} with a reduced bandwidth, as we show in Fig.~\ref{fig:2}a. Therefore, we expect to have the same LL in this case as in graphene.

%\redmark{Luisa, pls, reorganize this paragraph with the following flow of ideas:
%\begin{itemize}
%    \item In fig 2b we show the energy levels calculated with the complete/reduced TB.  for a ZZ  ribbon. \bf{Maybe you can choose a different value of B, so that the 2 sets of LL, one per valley, are clearly seen}. Add in second sentence the technical details of the number of states, $W=$ etc, as it is right now It is identical to the ones obtained in graphene (and cite a paper), except for the scaling of the total energy
%    \item In fig 2c the energy spectrum of fig 2a at $k=?$ and compare with the LL obtained with the kp theory obtained analytically (reference). We find very good agreement, specially at low energies, where kp theory works better.
%    \item  Discuss briefly that the TB bands have flat bands, whose wave functions are located away from the ribbon edge, and dispersive bands associated to edge states 
%\end{itemize}}

In Fig.~\ref{fig:2}b we show the energy levels for $[2,2]$triangulene zigzag ribbons under the influence of a magnetic field. These were obtained by numerical diagonalization of the reduced TB Hamiltonian under  $B=6$ T with $N_{td}=100$ triangulene dimers and width $W_{[2,2]}=524$ nm. The full tight-binding model yields $N_t=26N_{td}$ energy bands, wherein $(n_a+n_b-2)N_{td}=2N_{td}$ bands are located around the Fermi energy inside a large gap. This latter set of bands is well described using the reduced TB Hamiltonian on the basis of the two zero modes per triangulene dimer, where every $[2]$triangulene hosts exactly one isolated zero mode. The low-energy sector is found to be isomorphic to the tight-binding model describing the $p_z$ orbitals of graphene on a zigzag ribbon \cite{Lado2015} with effective hopping term of $\tilde{t}_g=\frac{t_3}{3}$ and energy bandwidth of $6\tilde{t}_g$ (see Fig.~\ref{fig:2}a,b).

The energy levels centered around K from numerical diagonalization as well as $k\,\cdot\,p$ theory are plotted in Fig.~\ref{fig:2}c. 
The $k\,\cdot\,p$ Hamiltonian is isomorphic to that of graphene and reads
\begin{equation}
 	H_n^{[2,2]}=
    %A_0
    \varepsilon^{[2,2]}_0 \begin{pmatrix}
		0 & a^\dagger\\
		a& 0
        \label{kp:22}
	\end{pmatrix}
\end{equation} 	
We find very good agreement, specially at low energies, where $k\,\cdot\,p$ approximation holds. Both approaches  reveal the well-known graphene LL scaling \cite{Novoselov2005} of 
\begin{equation}
    \frac{E_n^{[2,2]}}{\varepsilon_0^{[2,2]}}=\pm\sqrt{n+1}
    \label{eq:LL22}
\end{equation}
with $n$ a non-negative integer and $\varepsilon_0^{[2,2]}=5t_3a_{CC}/\sqrt{2}\ell_B$. 
In addition, the Hamiltonian of Eq.~(\ref{kp:22}) has a zero-energy state for each $k_x$, also obtained in the tight-binding calculation.
For $B=10$~T, $t_3=0.1 t\simeq 270$ meV, $a_{CC}=1.42$~\AA,  we obtain $\ell_B=81$~\AA~and $\varepsilon_0=17$ meV. The analytical Eq.~\eqref{eq:LL22} is supported by the Dirac-Landau spectrum obtained analytically from the effective mass description. The magnitude $\varepsilon_0^{[2,2]}$ being approximately six times lower than that of graphene serves as validation of our numerical results once the relation between the Fermi velocities of the Dirac fermions for the two models is known to be $v_F^{[2,2]}=v_F^G/6$. Our results confirm that the low-energy Landau spectrum  of the $[2,2]$triangulene crystal reproduces that of graphene, with an energy scale reduced by a factor of six for the parameters used here.

\begin{figure*}
    \centering
    \includegraphics[width=1\textwidth]{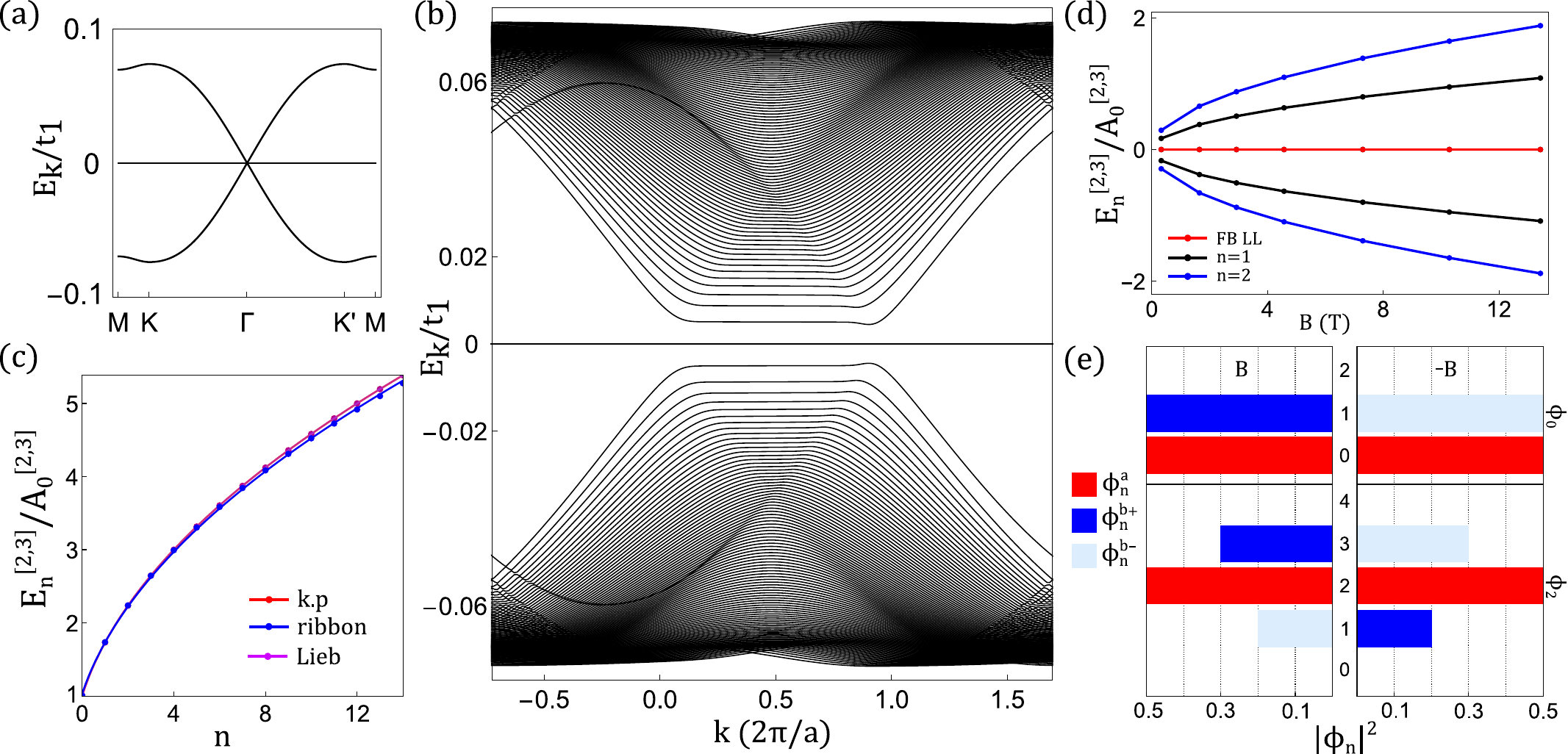}
    \caption{Landau level structure of $[2,3]$triangulenes. (a) Energy bands of the crystal obtained using the reduced TB model, with $t_3 = 0.1t_1$. (b) Energy bands of the zigzag ribbon configuration with $N_{td}=100$ dimers in the unit cell and magnetic field set to $B=11.4$ T. The results feature
    two sets of Landau levels in the vicinity of the $\Gamma$-point. (c) Scaling of the first Landau levels show the $\sqrt{2n+1}$ behavior both in the numerics (blue) and using $k\,\cdot\,p$ theory (red), analogous to the scaling found in the Lieb model (purple). (d) Behavior of the first Landau levels with external field ($B$) show no LL spreading of the $E=0$ manifold with $N_{td}$ subbands. (e) Orbital components for the lowest positive energy Landau state ($n=0$) (top) and $n=2$ state (bottom) in the vicinity of the $\Gamma$-point for two field directions (left/right).}
    \label{fig:3}
\end{figure*}

We note that we find two types of TB magnetic bands in the numerical diagonalization shown in Fig.\ref{fig:2}b. Flat levels that make up the Landau level spectrum arise from bulk localization of the wave functions away from the ribbon edges. On the other hand, dispersive bands are associated to magnetic edge states located in the open boundaries. We also note that, at $E=0$, there are both \cite{Brey2006,Lado2013} bulk Landau levels and dispersionless edge states, already  present \cite{Nakada1996} at $B=0$, degenerate with the zero energy LL.

\section{Landau levels in $[2,3]$triangulene \label{23}}

\subsection{Tight-binding results}

We now calculate the Landau levels of the $[2,3]$triangulene crystal. 
The low-energy bands of this system \cite{Ortiz2022} shown in Fig.~\ref{fig:3}a are isomorphic to the Lieb lattice, with a pair of bands forming a Dirac cone at the $\Gamma$-point and a $E=0$ band degenerate with the other pair at the $\Gamma$-point.  The zero-energy manifold persists at finite $B$ because the Peierls phases preserve the bipartite structure and sublattice imbalance, although the corresponding wave functions depend on the magnetic field.
Therefore, at finite $B$, still one third of the macroscopic number of low-energy states associated with zero modes remains at $E=0$ for every $k_x$, identical to the $B=0$ limit. This behavior differs from the $[2,2]$triangulene case, where only a single zero-energy state per $k_x$ is present within each valley at finite $B$.  

In Fig.~\ref{fig:3} we show the low-energy bands $E_n(k_x)$ of zigzag $[2,3]$triangulene ribbons with $N_{td}=100$ unit cells that corresponds to a width of $W_{[2,3]}=629$ nm, calculated using the reduced tight-binding model with magnetic field set to $B=11.4$ T. The total number of bands is therefore $3N_{td}$. These low-energy bands are identical to those obtained using the full TB Hamiltonian, confirming that the wave functions of the low-energy bands are formed almost exclusively by the zero modes of the monomers.

We identify a set of flat $N_{td}$-fold degenerate $E=0$ bands across the full Brillouin zone and a set of flat bands in a wide $k$-region around the $\Gamma$-point that become dispersive otherwise (see Fig.~\ref{fig:3}b). These $E=0$ states are also obtained \cite{Madail2024} in the absence of magnetic field and are warranted to appear on account of the single-particle limit of Lieb  theorem\cite{Lieb1989} that relates the minimal number of zero-energy states to the sublattice imbalance of the unit cell. In the present case, there is a sublattice imbalance of $N_{td}$. Hence, we conclude that the $E=0$ states are not induced by the magnetic field. %Therefore, unlike the case of LL for Dirac electrons, there is no $E=0$ LL for the Dirac cone of the  $[2,3]$triangulene crystal. We also note that the zero modes associated to the $E=0$ flat band of this lattice are localized in one sublattice and  their energies are not shifted by the magnetic field, on account of the absence of orbital moment.

The two sets of electron-hole symmetric finite-energy flat bands in Fig.~\ref{fig:3}b correspond to the Landau levels associated with the linear dispersive bands of the 2D crystal. The same bands become dispersive at different values of $k_x$ away from $\Gamma$-point.
An analysis of the corresponding wave functions indicates that the flat $k$-region correspond to LL states with bulk localization while becoming magnetic edge states away from the center of the Brillouin zone.   

\subsection{$k\,\cdot\,p$ results}
We now derive the expressions for the Landau level scaling using the $k\,\cdot\,p$ method. Interestingly, this can be derived analytically and verified against numerical tight-binding results (see 
Fig.~\ref{fig:3}c). 
We obtain the following $k\,\cdot\,p$ Hamiltonian after expansion around the $\Gamma$-point and Peierls substitution

\begin{equation}
H_n^{[2,3]}=\epsilon_0^{[2,3]}\begin{pmatrix}
    0&a^\dagger&-a\\
    a&0&0\\
    -a^\dagger&0&0
    \label{Hn23}
\end{pmatrix}    
\end{equation}
with $\epsilon_0^{[2,3]}=6t_3a_{CC}\sqrt{\frac{3}{11}}\frac{1}{\ell_B}$. Note that Eq.~(\ref{Hn23}) contains only linear terms in $a$ given that we took the first order expansion around the $\Gamma $ point,  where the bands have a linear dispersion. We write the Schrödinger equation
\begin{equation}
    H_n^{[2,3]}\ket{\Phi_n^{[2,3]}}=E_n^{[2,3]}\ket{\Phi_n^{[2,3]}},
\end{equation}
and we introduce the notation
for an eigenstate $\ket{\Phi_n}=\left(\ket{\Phi_n^A},\ket{\Phi_n^{B_1}},\ket{\Phi_n^{B_2}}\right)^T$, where each component represents the projection of the Landau level wave function over one of the three zero modes in the unit cell. The Schr\"odinger equation can be represented as the
linear system
\begin{eqnarray}
\epsilon_0^{[2,3]}(a^\dagger\ket{\Phi_n^{B_1}}-a \ket{\Phi_n^{B_2}})&=E_n^{[2,3]}\ket{\Phi_n^{A}},\\
\label{eigsys1}
\epsilon_0^{[2,3]}a \ket{\Phi_n^{A}}&=E_n^{[2,3]}\ket{\Phi_n^{B_1}},\\
\label{eigsys2}  
-\epsilon_0^{[2,3]}a^\dagger\ket{\Phi_n^{A}}&=E_n^{[2,3]}\ket{\Phi_n^{B_2}}.
\label{eigsys3}
\end{eqnarray}
The space of $3N$ solutions of this operator is divided in three groups with dimension $N$. Two groups of electron-hole symmetric states with finite energy $\pm E_n$, and a group with $E_n=0$. 

\subsubsection{Finite-energy solutions}
In order to find the finite-energy solutions we substitute
the two last equations for $\ket{\Phi_n^{B_1}}$ and $\ket{\Phi_n^{B_2}}$ onto the first equation and obtain the closed form for $\ket{\Phi_n^{A}}$ in 
\begin{equation}
\epsilon_0^{[2,3]^2}(a^\dagger a+aa^\dagger)\ket{\phi_n^A}=
E_n^{[2,3]^2} \ket{\phi_n^A}.
    \label{eq:23an}
\end{equation}
It is apparent that
%The energy dispersion can be retrieved by assuming the Fock state in A-orbital 
$\ket{\Phi_n^A}=\ket{n}$ solves Eq.~\eqref{eq:23an} with 
\begin{equation}
    E_n^{[2,3]}=\pm \epsilon_0^{[2,3]}\sqrt{2n+1},
    \label{eq:23LLkp}
\end{equation} 
where $n$ is a non-negative integer. Comparison between Eq.~\eqref{eq:23LLkp} and our numerical results for the
reduced model show a very good agreement 
(see Fig.~\ref{fig:3}c). 
Using $\ket{\Phi_n^A}\propto\ket{n}$ obtained from Eq.~\eqref{eq:23an}, we can now use Eq.~(\ref{eigsys1}) to obtain the  
  %having found the solution for $$, it is straightforward to obtain 
  the other components of the wave function. For $B>0$, we find
\begin{equation}
   \ket{\Phi_0^{[2,3]}}=\tfrac{1}{\sqrt{2}}\begin{pmatrix}
       \ket{0}\\
       0\\
       \mp\ket{1}
    \end{pmatrix},\,\,\ket{\Phi_n^{[2,3]}}=\tfrac{1}{\sqrt{2}}\begin{pmatrix}
       \ket{n}\\
       \pm\chi_n\ket{n-1}\\
       \mp\xi_n\ket{n+1}
    \end{pmatrix},
    \label{eq:23kpfinite}
\end{equation}
with $n>0$, $\chi_n\equiv \frac{\sqrt{n}}{\sqrt{2n+1}}$ and $\xi_n\equiv \frac{\sqrt{n+1}}{\sqrt{2n+1}}$ . For $B<0$, the role of the components $B_1$ and $B_2$ is swapped. Each component of the finite-energy eigenvectors in Eq.~\eqref{eq:23kpfinite} is tied to the single Landau index $n$. The general construction already incorporates the choice of the vacuum on the $B_1$-orbital (for $n=1$) and taking $\ket{\Phi_0^{B_2}}\equiv\ket{0}$ is not a valid solution, since it fails to satisfy the second equation of the coupled system. Thus, the extra $n=0$ doublet is found by setting the $A$-orbital component $\ket{\Phi_0^{A}}\equiv\ket{0}$.  As an illustration, states $\Phi_0$ and $\Phi_2$, found from the numerical diagonalization of the effective-mass Hamiltonian of Eq.~(\ref{Hn23}), are shown in Fig.~\ref{fig:3}e. We find that the three orbital components follow our analytical calculation. Moreover, reversing the direction of the external field transforms the ladder operators as $a\rightarrow -a^\dagger$ and the states undergo orbital exchange.
Indeed, the Hamiltonian in the reversed field is equivalent to
\begin{equation}
    H_n^{[2,3]}(-B)=O\,H_n^{[2,3]}(B)\,O^{-1},
\end{equation}
where we define the orbital exchange operator $\hat{O}$ from the transformation $\big(\ket{\Phi_n^A},\ket{\Phi_n^{B_1}},\ket{\Phi_n^{B_2}}\big)\rightarrow\big(\ket{\Phi_n^A},\ket{\Phi_n^{B_2}},\ket{\Phi_n^{B_1}}\big)$. This result implies a swap in the orbital character when flipping the field direction and can be understood as follows. The representation of the reduced TB model relies on the $B=0$ zero modes of the triangulene as eigenstates of the $C_3$ operator.  With this choice, the zero mode wave functions are complex conjugates of each other \textit{i.e.} they relate through a change of sign of the local phase \cite{Ortiz2022,Madail2024}. Both the change of sign of magnetic field, accounted for in the Peierls substitution, and the orbital swap imply a change of sign of the phase of the zero modes.

\subsubsection{Zero-energy solutions}

Due to the presence of the crystallographic zero-energy flat band in $[2,3]$triangulene, a $E=0$ magnetic manifold is also expected.   
In this case, the last two equations of the system collapse to $a \ket{\mathcal{Z}_n^A} = a^\dagger \ket{\mathcal{Z}_n^A} = 0$, whose only solution is the trivial $\ket{\mathcal{Z}_n^A} = 0$. The bipartite property of the lattice is evident in the sublattice polarization of the zero-energy states - every zero-energy state must have vanishing weight on the $A$-orbital. The remaining equation of the system yields
\begin{equation}\label{230constraint}
a^\dagger\ket{\mathcal{Z}_{0}^{B_1}} = a\ket{\mathcal{Z}_{0}^{B_2}}.
\end{equation}
the operator relation coupling the two $B$-sector components. 
Unlike the finite-energy case, there is no eigenvalue constraint on a single Landau index. Consequently,
any superposition of $\ket{n}$ levels on the $B$-orbitals that satisfies the ladder-operator relation is allowed.  Eq.~(\ref{230constraint}) may be satisfied by following a solution of the form
\begin{equation}
\ket{\mathcal{Z}_n^{B_1}} = b_1 \ket{n-1}, \qquad
\ket{\mathcal{Z}_n^{B_2}} = b_2 \ket{n+1},
\end{equation}
where $b_1$ and $b_2$ are coefficients that are found by requiring that
Eq. (\ref{230constraint}) is satisfied, ie.,  $ \sqrt{n}\, b_1 = \sqrt{n+1}\, b_2 $, and normalization condition $b_1^2+b_2^2=1$. Hence, a particularly simple choice for the $E=0$ wave functions reads
\begin{equation}
\ket{\mathcal{Z}_0^{[2,3]}} = \begin{pmatrix} 0\\0\\ \ket{0}
\end{pmatrix},\;
    \ket{\mathcal{Z}_n^{[2,3]}}=\tfrac{1}{\sqrt{2n+1}}
\begin{pmatrix}
0 \\
\sqrt{n+1}\ket{n-1} \\
\sqrt{n}\ket{n+1}
\end{pmatrix}.
\end{equation}
The $n=0$ case is found by assuming solutions of the form $\ket{\mathcal{Z}_n^{B_2}} \equiv \ket{0}$ which gives the relation between B-orbitals $    a^\dagger \ket{\mathcal{Z}_n^{B_1}}=a \ket{\mathcal{Z}_n^{B_2}}=0 $ and is only true if $ \ket{\mathcal{Z}_n^{B_1}}=0$. 
The zero-energy eigenstates are extended over many Landau levels along the $B$-orbitals reflecting the flat band nature of the manifold.

\begin{figure*}
    \centering
    \includegraphics[width=1\textwidth]{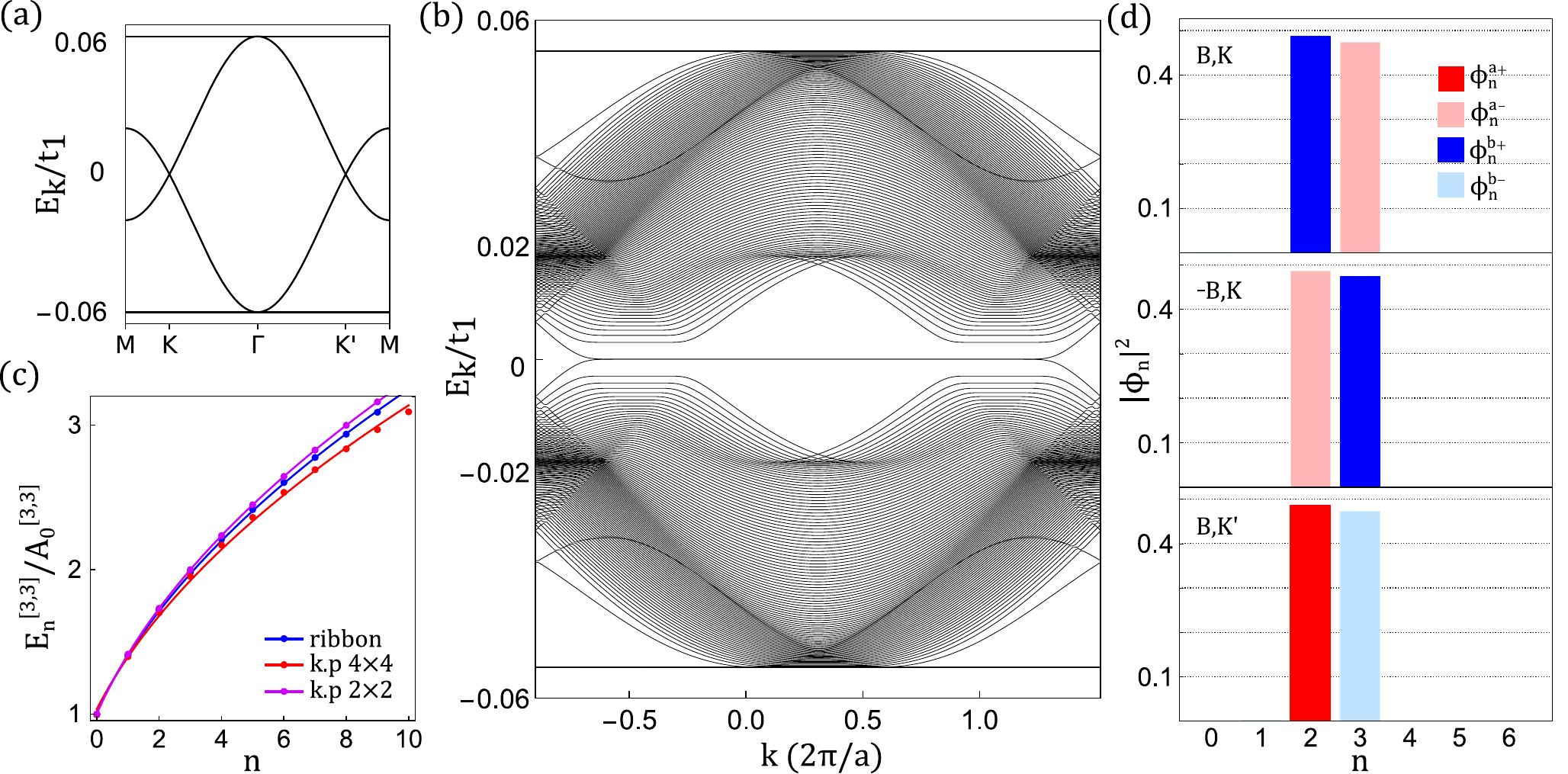}
    \caption{Landau level structure of $[3,3]$triangulenes. (a) Energy bands of the crystal obtained using the reduced TB model, with $t_3 = 0.1t_1$. (b) Energy bands of the zigzag ribbon configuration with $N_{td}=100$ dimers in the unit cell and magnetic field set to $B=4\,\rm T$. The results show Landau level structure centered around each Dirac valley and in the vicinity of the quadratic band-crossing points around the $\Gamma$-point. (c) Scaling of the first Dirac-Landau levels show the $\sqrt{n+1}$ behavior both in the numerics (blue) and using $k\,\cdot\,p$ theory on the full Hamiltonian (red) and on the reduced block (purple). (d) Orbital components of the $n=2$ eigenstate of the full Landau Hamiltonian at the $K$ (top) and $K'$ (bottom) valleys and at a reversed external field (middle).}
    \label{fig:4}
\end{figure*}

\subsection{Comparison with the Landau Levels of  the Lieb lattice}
Given the fact that the bands of the [2,3] triangulene lattice are isomorphic to those of the Lieb lattice, except for the fact that in the Lieb lattice triple-degeneracy point is at the $M$ point instead of the $\Gamma$ point, we now compare the Landau levels of both systems. Under the influence of a magnetic field, the dispersive energy bands of the Lieb lattice  follow the conventional Onsager rule \cite{Jakubsky2022,Rhim2020}. As expected, the non-dispersive band, whose $B=0$ states are localized in the majority sublattice and have no current and therefore no orbital moment, are not affected by the magnetic field. 

The $k\,\cdot\,p$ Hamiltonian for the Lieb lattice (see Appendix~\ref{SMB}) reads 
\begin{equation}
H_n^{\text{Lieb}}=\frac{\epsilon_0^{\text{Lieb}}}{\sqrt{2}}\begin{pmatrix}
    0&-i(a+a^\dagger)&a-a^\dagger\\
    i(a+a^\dagger)&0&0\\
    a^\dagger-a&0&0
    \label{HnLieb}
\end{pmatrix}    
\end{equation}
with $\epsilon_0^{\text{Lieb}}=a_Lt/\ell_B$. In contrast with the case in Eq.~(\ref{Hn23}), the off-diagonal terms of Eq.~(\ref{HnLieb}) now depend on both ladder operators. 
The analytical solution of Eq.~(\ref{HnLieb}) shows that, on top of the $E=0$ band, the LL spectrum of the Lieb lattice follows the same rule as the $[2,3]$triangulene crystal of 
$E_n=\epsilon_0^{\text{Lieb}}\sqrt{2n+1}$ [see Eq.~\eqref{eq:23LLkp}]. This result can also be retrieved from the Onsager relation \cite{Onsager1952,Mikitik1999} for a linear dispersive band  with band velocity $v_L$
$S(E)=\frac{\pi E^2}{(\hbar v_L)^2}
=\frac{2\pi e|B|}{\hbar}
\left(n+\frac{1}{2}-\frac{\gamma_B}{2\pi}\right)$. For the Lieb lattice, the Berry phase is $\gamma_B=0$ and the LL quantization is found to be proportional to $\sqrt{2n+1}$. The normalized Landau states are written as 
\begin{equation}
\ket{\Phi_n^{\mathrm{Lieb},\pm}} =
\dfrac{1}{\mathcal{N}_{\rm Lieb}}
\begin{pmatrix}
\sqrt{4n+2}\,\ket{n} \\[2pt]
\pm i\left(
\sqrt{n}\,\ket{n-1}
+\sqrt{n+1}\,\ket{n+1}
\right) \\[2pt]
\pm\left(
\sqrt{n+1}\,\ket{n+1}
-\sqrt{n}\,\ket{n-1}
\right)
\end{pmatrix}
\end{equation}
for $n\geq0$ with $\mathcal{N}_{\rm Lieb}=\sqrt{8n+4}$. It is apparent by inspection of the analytical solutions that the finite-energy Landau states of the Lieb lattice are analogous to those of the $[2,3]$triangulene, up to a basis rotation among the $B$ and $C$ orbitals. The $E=0$ manifold is also present in the Lieb lattice and the corresponding solutions are equivalent to B-orbitals hybridization of the zero-energy states of the $[2,3]$triangulene.
We therefore conclude that
the two-dimensional $[2,3]$triangulene is the  $C_3$ symmetric version of the square two-dimensional Lieb lattice,
with the same low-energy bands and the same LL spectrum.

\section{Landau levels in $[3,3]$triangulene \label{33}}

The two-dimensional $[3,3]$triangulene crystal features two Dirac cones and two band crossings at the $\Gamma$-point \cite{Ortiz2022} at energies 
$E_{\rm FB}^{(\pm)}=\pm 3 t_\text{eff}^{[3,3]}=\pm 6t_3/11$  (see Fig.~\ref{fig:4}a).  Naively, we expect that the set of Landau levels associated to the $E\sim 0$ states at the $K$ and $K'$ points should be very similar to those of graphene. Below, we show that both the tight-binding and $k\cdot p$ calculations confirm this scenario,  but we also find some deviations whose origin is discussed. On the other hand, the band crossings at the $\Gamma$-point are reminiscent of a singular flat band \cite{Rhim2019}.
At these quadratic band crossings, conventional Landau levels originating from the dispersive band coexist with an anomalous sequence associated with the singular flat band.
%In a singular flat band, two sets of LL are expected: a set that follows the standard Dirac-Landau scaling of the energy levels and a second with a very peculiar $1/n$ scaling. 

\subsection{Landau levels associated to the Dirac cones}

In Fig.~\ref{fig:4} we show the low-energy bands $E_n(k_x)$ computed with the reduced TB model for a zigzag ribbon with $W_{[3,3]}=733$ nm  and $B=4T$.
The choice of magnetic field $B=4$ T ($\ell_B=128$ \AA) satisfies
the required separation of length scales,
$a_{CC} \ll \ell_B \ll W_{[3,3]}$.
Thus, the magnetic length is much larger than the microscopic
lattice scale, as required for the $k\cdot p$ description, while remaining much smaller than the ribbon width, allowing for well-localized bulk Landau states.

%We are interested in capturing the effect of the magnetic field in both of these points, from the low-energy bands $E_n(k_x)$ of zigzag ribbons with $W_{[3,3]}=733$ nm in Fig.~\ref{fig:4}. 
The  four $B=0$ two-dimensional bands (see Fig.~\ref{fig:4}a)  split, under the influence of the magnetic field,  into the 4$N_{td}$ sub-bands of Fig.~\ref{fig:4}b.
There is full agreement between these low-energy bands extracted from the reduced TB model and those obtained using the full TB Hamiltonian, except at the energy window of the flat bands. The reduced model does not reproduce the small energy broadening of these bands seen in the full calculation even at zero field~\cite{Madail2024}. The origin of this effect comes from hybridization of higher/lower energy states that are missing in the reduced TB model.

\subsubsection{Dirac-Landau spectrum and validity of the Dirac approximation}

In Fig.~4(b) we identify a sequence of nearly dispersionless magnetic bands around the two valleys at half filling. These are the bulk Landau levels originating from the Dirac cones of the two-dimensional crystal. Away from the valleys, they evolve into dispersive magnetic edge states localized at opposite boundaries
of the ribbon according to the sign of their group velocity.

The low-energy Landau spectrum can be understood by expanding the reduced $[3,3]$ Bloch Hamiltonian around either valley and
projecting onto the two-dimensional zero-energy manifold. At the $K$ valley this subspace is spanned by the $a_-$ and $b_+$ zero
modes, whereas at $K'$ it is spanned by the complementary $a_+$ and $b_-$ modes. To leading order in momentum, the projected Hamiltonian is a two-component Dirac Hamiltonian.
After minimal coupling, its spectrum is
\begin{equation}
E^{[3,3]}_{K,n}
=
\pm \epsilon^{[3,3]}_0\sqrt{n+1},
\qquad n=0,1,\ldots ,
\label{eq:33_Dirac_LL}
\end{equation}
with $\epsilon^{[3,3]}_0
=
\frac{21\sqrt{2}}{11}
\frac{a_{\rm CC}}{\ell_B}t_3$.
Thus, sufficiently close to the Dirac point, the $[3,3]$triangulene crystal displays the same $\sqrt{B}$ and
$\sqrt{n+1}$ Landau-level scaling as graphene, with a
renormalized velocity.

However, as shown in Fig.~4(c), this Dirac approximation becomes
progressively less accurate with increasing Landau-level index.
The deviations are already reproduced by the $k\cdot p$
Hamiltonian and therefore do not originate from lattice-scale or
edge effects. Their origin is instead the multiband character of
the $[3,3]$ Dirac fermions. The two zero-energy states coexist
with a pair of states at finite energies $\pm E_{\rm FB}$ and the
linear-in-momentum Hamiltonian couples the Dirac and remote
sectors. In the eigenbasis at $K$, the $k\cdot p$ Hamiltonian can
be written schematically as
\begin{equation}
H_{kp}'=
\begin{pmatrix}
H_{\rm D} & V\\
V^\dagger & H_{\rm FB}
\end{pmatrix},
\label{eq:33_block}
\end{equation}
where $H_{\rm D}$ is the two-component Dirac Hamiltonian,
$H_{\rm FB}$ describes the remote finite-energy sector, and $V$
couples the two manifolds. Their explicit forms are given in
Appendix~\ref{app:33K}.

For a Landau state of index $n$, the relevant matrix elements of
$V$ contain ladder operator factors and therefore scale as
\begin{equation}
|V_n|\sim
\epsilon^{[3,3]}_0\sqrt{n+1}.
\end{equation}
Consequently, the validity of the isolated Dirac description is
controlled parametrically by
\begin{equation}
\lambda_n
\sim
\frac{|V_n|}
{|E_{\rm FB}-E_n^{\rm D}|}.
\label{eq:33_mixing_parameter}
\end{equation}
For low $n$ and weak magnetic field, $\lambda_n\ll1$ and the
$\sqrt{n+1}$ spectrum is recovered. Increasing either $n$ or
$B$ enhances the coupling to the remote sector, while the
energy separation from the finite-energy bands is reduced.
The resulting hybridization explains the deviations from the ideal Dirac-Landau sequence observed in Fig.~4(c).

\subsubsection{Properties of the Landau level wave functions}

We now consider the internal structure of the Landau-level
wave functions. As discussed above, the low-energy Dirac
manifold at the $K$ valley is spanned by the $a_-$ and $b_+$
states. In this basis, the projected Hamiltonian in a
perpendicular magnetic field takes the form
\begin{equation}
H_{K}^{\rm D}(B)
=
-\epsilon^{[3,3]}_0
\begin{pmatrix}
0 & a^\dagger\\
a & 0
\end{pmatrix},
\label{eq:33_Dirac_B}
\end{equation}
Accordingly, the corresponding eigenstates are formed by harmonic oscillator states whose Landau indices
differ by one.
Embedding these two-component eigenstates into the
original basis ${a_+,a_-,b_+,b_-}$ gives
\begin{equation}
\Phi^{[3,3]}_{K,n,\pm}
=
\frac{1}{\sqrt{2}}
\begin{pmatrix}
0\\
\ket{n+1}\\
\mp \ket{n}\\
0
\end{pmatrix},
\qquad n=0,1,\ldots .
\label{eq:33LLWF}
\end{equation}

The numerical solution of
the full $k\cdot p$ Hamiltonian in Eq.~\eqref{eq:33_block} is consistent with Eq.~\eqref{eq:33LLWF}. In Fig.~\ref{fig:4}d we plot the components of the $n=2$ state. The small imbalance in the components of the full Hamiltonian solutions arise because of the coupling to the flat band sector through ladder operators with different matrix elements, $\sqrt{n+2}$ and $\sqrt{n}$, producing unequal leakage from the two active Dirac components. For higher values of $n$, the mixing with the flat band subspaces breaks this perfect valley-dependent sublattice-orbital locking. 

We now consider how the wave functions of Eq.~\eqref{eq:33LLWF} transform upon valley and magnetic field inversion. 
The two operations act differently on the internal structure of the Dirac-Landau spinor.
Under valley inversion, the active orbital pair changes from
$(a_-,b_+)$ to $(a_+,b_-)$, while the oscillator indices
associated with the two sublattices are exchanged
\begin{equation}
\Phi_{K',n,\pm}^{[3,3]}(B)
    =
    \frac{1}{\sqrt{2}}
    \begin{pmatrix}
        \pm\ket{n}\\
        0 \\
        0\\
         \ket{n+1}
    \end{pmatrix},
    \qquad n > 0.
\label{eq:33LLWFK'}
\end{equation}

By contrast, the state under $-\mathbf{B}$ remains within the same low-energy doublet, but its two active components are exchanged in both sublattice and orbital character

\begin{equation}
    \Phi_{K,n,\pm}^{[3,3]}(-B)
    =
    \frac{1}{\sqrt{2}}
    \begin{pmatrix}
        0\\
        \ket{n}\\
        \pm \ket{n+1}\\
        0
    \end{pmatrix},    
    \qquad n > 0.
\label{eq:33LLWFBneg}
\end{equation}
This distinction highlights the multiorbital nature of the Dirac manifold.
In graphene, the low-energy Hilbert space contains only one orbital per site, so the Landau states are fully described by the sublattice pseudospin. As a result, valley inversion and magnetic-field reversal act equivalently on the Dirac sector $H_{K}(-B)=H_{K'}(B)$. In the present case, this equivalence no longer holds because the additional orbital degree of freedom participates in the transformation under field reversal. Orbital textures in valley eigenstates are a characteristic feature of multiorbital models such as the $p_x,p_y$ honeycomb lattice \cite{Congjun2008,Canonico19}. Our Landau-level analysis shows that the same physics survives in the present zero-mode realization. At fixed valley, the low-index Dirac spinor is a coherent superposition of the two active orbital-sublattice components. The two valleys are in turn associated with orthogonal internal sectors, so that a coherent superposition of valley states necessarily combines the valley degree of freedom with a corresponding change in this internal texture.

%In the ribbon TB calculations, an additional zero-energy state can be found coexisting with the $n=0$ Landau level at each valley. This state corresponds to a bearded edge solution present in $[3,3]$triangulene ribbons. Without magnetic field, this twice degenerate state has zero-energy and is localized at the edge monomers when $0<|k|<2\pi/3$. For $B\neq0$, the region in k-space where this edge state survives expands with increasing magnetic field. 

\begin{figure}[t]
    \centering
    \includegraphics[width=1\linewidth]{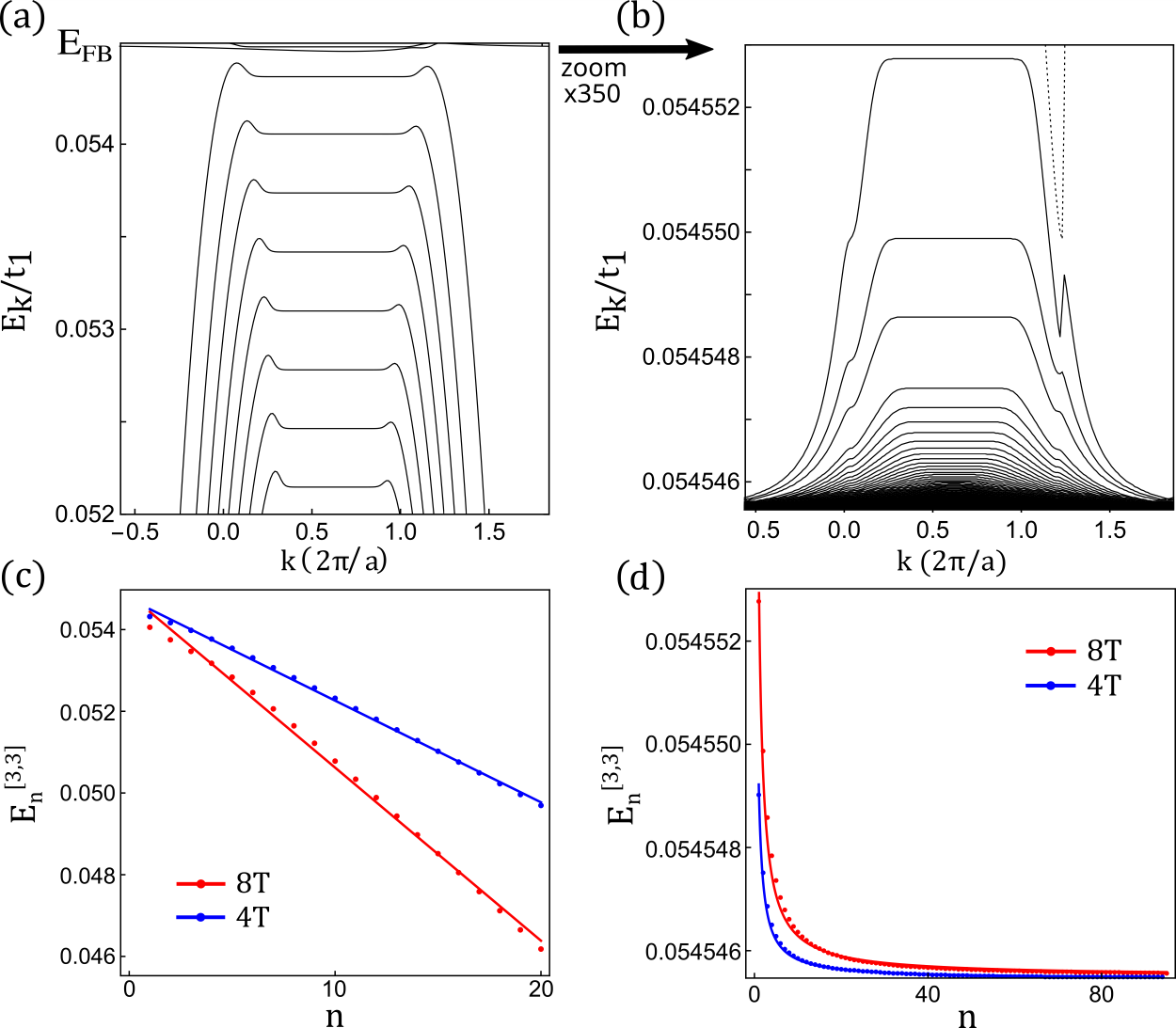}
    \caption{Landau quantization of the $[3,3]$triangulene at the positive-energy quadratic band crossing point. (a) Close-up on the LL of the dispersive band and (b) zoom on the LL of the flat band for $B=8$~T. The dashed curve belongs to an edge state present at zero field. (c) Linear scaling of the dispersive band and (d) inverse-law scaling of the flat band, for $B=8$~T (red) and $B=4$~T (blue).}
    \label{fig:5}
\end{figure}

\subsection{Landau levels associated to the singular flat band }
At the $\Gamma$ point there are two quadratic band crossings, where the  dispersive bands become degenerate with the flat bands at $E_{FB}^{(\pm)}=\pm E_{FB}$.
Recent work in systems with such band crossing points \cite{Rhim2020,Fu2025,Long2025} has reported the presence of Landau level splitting with anomalous behavior, specifically, with a $1/n$ scaling \cite{Rhim2019,Rhim2020,Oh2022}, together with a set of LL with conventional scaling, linear in $n$.
This anomalous LL scaling is thought to be the result of level repulsion between the Landau levels developed from the flat and the parabolic bands. In such cases, the flat band is singular and can be characterized by a nonzero Hilbert-Schmidt quantum distance 
\begin{equation}
d^2(\mathbf{k},\mathbf{k}')
=1-\left|
\braket{\psi_{\mathrm{FB}}(\mathbf{k})|
\psi_{\mathrm{FB}}(\mathbf{k}')}
\right|^2
\end{equation}
of the flat band wave function at $k$ and $k'$ sufficiently close to the singular point. The strength of the singularity, determined by the maximum value of $d\in[0,1]$, is linked to the total spread of the flat band  anomalous Landau levels \cite{Rhim2019,Rhim2020}. 

We have computed the quantum distance of the flat band Bloch eigenstates from the zero-field $[3,3]$ Hamiltonian close to the $\Gamma$-point crossing. We parametrize the momentum as
$\mathbf{q}=q(\cos\theta,\sin\theta)$ and consider the limit $q\rightarrow0$, with $\theta$ defining the direction of approach to the singular point. The quantum distance follows
\begin{equation}
    d^2(\theta_1,\theta_2)
    =
    \sin^2(\theta_1-\theta_2),
\label{eq:42}
\end{equation}
and is maximal for two directions separated by $\pi/2$, yielding
\begin{equation}
    d_{\text{max}}^{[3,3]}=1,
    \qquad
    |\theta_1-\theta_2|=\frac{\pi}{2}.
    \label{eq:qdistance33}
\end{equation}
This result is consistent with previous calculations on the analytical flat band wave functions of the isomorphic $p_{x,y}$-honeycomb lattice \cite{Xu2022}.

%%%%%%%%%

We now use our reduced tight-binding description to determine if the same LL phenomenology reported before \cite{Rhim2019,Rhim2020,Oh2022} is found in our system.
We consider the same ribbon
geometry used in Fig.~\ref{fig:4}. We  find four sets of LL associated to the $\Gamma$ point, two at $E_{FB}$ and their electron-hole symmetric partner  $-E_{FB}$.  It is enough to discuss the LL at the positive-energy band crossing, depicted in Fig.~\ref{fig:5}.

%The corresponding Hamiltonian at $\Gamma$ and the behavior of the Landau states can be solved analytically. In the appendix section we show the derivation of the LL associated to the singular flat band. We first obtain the effective $2\times2$ $k\cdot p$ Hamiltonian [Eq.~\eqref{eq:Heff33}] in Appendix~\ref{app:QBC_eff}  \ref{app:QBC_LL}using the second order Löwdin projection procedure described in Appendix~\ref{app:eff}.  From Peierls substitution we obtain the Hamiltonian in terms of the ladder operators [Eq.~\eqref{eq:H33B}] in Appendix~\ref{app:QBC_LL}, that allows for closed analytical expressions [Eq.~\eqref{eq:sfbkp33}]. 

%%%%
The corresponding Hamiltonian at $\Gamma$ and the behavior of the
Landau states can be solved analytically. We first obtain the $2\times2$ 
effective $k\cdot p$ Hamiltonian
[Eq.~\eqref{eq:Heff33}] in
Appendix~\ref{app:QBC_eff},
using the second-order L\"owdin projection procedure described in
Appendix~\ref{app:eff}. Applying the Peierls substitution yields
the Hamiltonian in terms of ladder operators
[Eq.~\eqref{eq:H33B}] in Appendix~\ref{app:QBC_LL}, where we derive the exact Landau level sequences and
their large-$n$ asymptotic forms. The anomalous sequence has the
asymptotic behavior given in Eq.~\eqref{eq:sfbkp33}. In the large-$n$ limit, the LL associated with the singular flat
band scale as
\begin{equation}
    E_{n,-}^{[3,3](+)}
    \simeq
    E_{\rm FB}^{(+)}
    +
    \frac{3\gamma}{4\ell_B^2}\frac{1}{n},
    \quad n\gg1,
    \label{Engamma}
\end{equation}
where $\gamma=147t_3a_{\rm CC}^2/44$. The corresponding sequence is shown in
Fig.~\ref{fig:5}b,d. At the positive-energy crossing, this anomalous set lies above and approaches $E_{\rm FB}$ as $n$ increases. We note that the energy window of Fig.~\ref{fig:5}b is $\approx20\mu$eV for $B=8$~T, indicating that the levels lie extremely close in energy.

The second set of LL reported in Fig.~\ref{fig:5}a,c behave as
\begin{equation}
    E_{n,+}^{[3,3](+)}
    \simeq
    E_{\rm FB}^{(+)}
    -
    \frac{4\gamma}{\ell_B^2}
    \left(
        n-\frac{1}{2}
    \right),
    \quad n\gg1,
    \label{EQBgamma}
\end{equation}
and have the conventional linear scaling proportional to
$-nB$. The energy separation in this case is $\approx 0.8$ meV, for $B=8$~T in the ribbon calculation, which is $40\times$ larger than the energy spread of the anomalous counterpart.

The results of this subsection show that the flat bands involved in the quadratic band crossings of the $[3,3]$triangulene exhibit the characteristic signatures of  singular flat bands \cite{Rhim2021} both at positive and negative energy, on account of the  electron-hole symmetry of bipartite lattices.

\begin{figure*}[!t]
    \centering
    \includegraphics[width=1\textwidth]{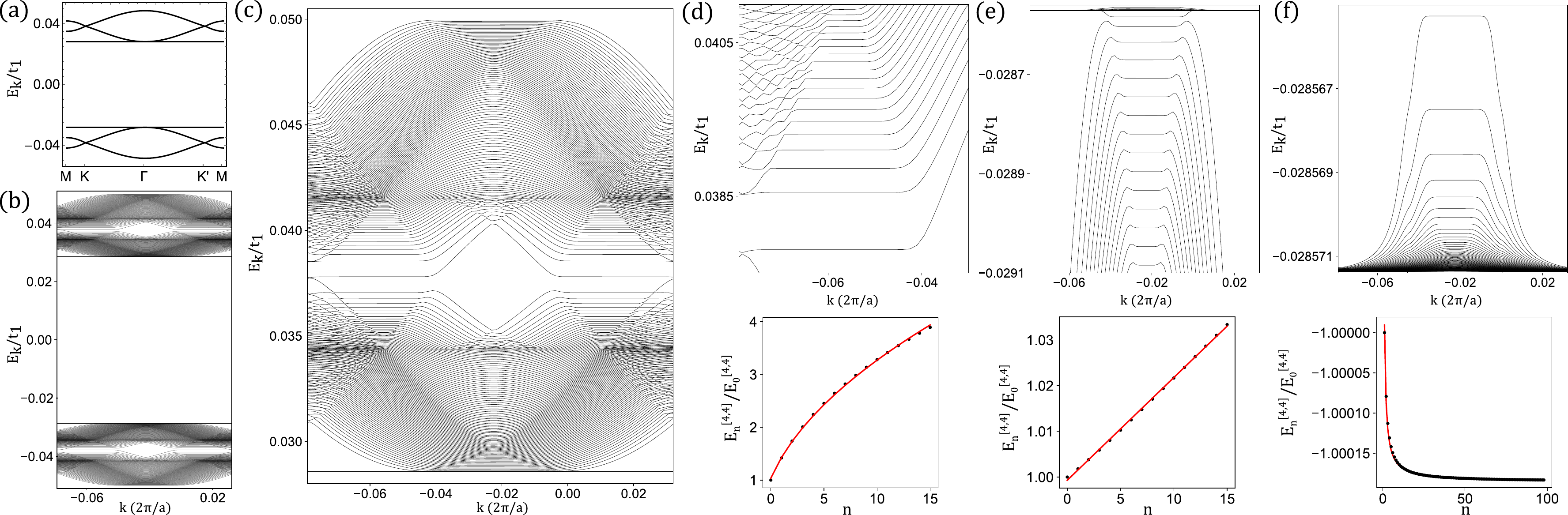}
    \caption{Landau level structure of $[4,4]$triangulenes. (a) Energy bands of the crystal obtained using the reduced TB model, with $t_3 = 0.1t_1$. (b) Energy bands of the zigzag ribbon configuration with $N_{td}=100$ dimers in the unit cell and magnetic field set to $B=4$ T and (c) close-up on the positive energy bands. Landau level structure can be seen around (d) the Dirac valley with a $\sqrt{n+1}$ scaling and in the vicinity of the quadratic band-crossing points with (e) a linear dependence originating from the quadratic band and (f) $1/n$ dependence from the flat band.}
    \label{fig:6}
\end{figure*}

\section{Landau levels in $[4,4]$triangulene \label{44}}
The  $B=0$ energy bands of the $[4,4]$triangulene, shown  in Fig.~\ref{fig:6}a, feature a gap at half-filling that separate two copies of Kagomé-like bands. %: a flat valence/conduction band that is degenerate at the $\Gamma$-point with the top/bottom Dirac bands. 
Each kagome-like sector contains a flat band that touches a dispersive band quadratically at $\Gamma$, while the dispersive bands form Dirac crossings at $K$ and $K'$
These bands can be found analytically \cite{Ortiz2022} through the basis set of  3 zero modes localized  the corners of each $[4]$triangulene. The effective hopping terms $\{\tilde{t},\eta\tilde{t}\}$ are obtained from first and second neighbor overlap between adjacent corner states, with $\eta$ the parameter controlling the dispersion of the bands. In a zigzag ribbon, the original bands split into 6$N_{td}$ sub-bands and two zero-energy edge states emerge in the middle of the bandgap. Upon application of an external magnetic field, we observe that both the bandgap and the zigzag edge states remain insensitive to the field induced symmetry breaking (see Fig.~\ref{fig:6}b,c), at the energy scale of the figure. Still, a closer look at the band structure reveals three different Landau spectra:
\begin{itemize}
    \item [(i)] Dirac-Landau levels with a square-root law scaling in the vicinity of the $K$ and $K'$ points (see Fig.~\ref{fig:6}d);
    \item [(ii)] Landau levels stemming from the dispersive band at the quadratic crossing point with an apparent linear behavior (see Fig.~\ref{fig:6}e);
    \item[(iii)] Anomalous Landau level scaling belonging to the flat band at the quadratic band crossing point (see Fig.~\ref{fig:6}f).
\end{itemize}

The momentum broadening of the plateaus for all LL sets, together with the
energy spread of the flat band LL, increases with the magnitude of the
magnetic field. %and is thus susceptible to field fluctuations
Nevertheless, the overall spread of the flat band LL remains approximately one to three orders of magnitude smaller than that of the other LL sets.

For the flat band eigenstates  of the $[4,4]$triangulene crystal in the neighborhood of $\Gamma$, we have carried out the calculation of the quantum distance and we obtain identical results to those of the $[3,3]$ crystal in Eqs.~\eqref{eq:42},\eqref{eq:qdistance33}. Therefore, we expect a similar set of LL \cite{Rhim2019,Rhim2020,Hwang2021} associated to the singular flat band at $\Gamma$.

%Unlike the $[3,3]$ case, where the anomalous LL were located above $E=E_{FB}$ and below $E=-E_{FB}$, in the $[4,4]$ case, both positive- and negative-energy anomalous Landau states penetrate the central insulating gap. We note that recent theory work has shown that anomalous Landau levels extending into an insulating gap can produce oscillatory magnetization and conductivity even in the absence of a zero-field Fermi surface, suggesting that [4,4]triangulenes may likewise exhibit quantum oscillatory signatures \cite{Fu2025,Panda2022,Shao2024}.

%In this section, 
We now derive the scaling behavior of the anomalous Landau spectrum. 
%Because the relevant subspace is not centered at zero-energy, we are not able to recover this behavior from the full Hamiltonian $k\,.\,p$ approximation up to second order in $k$. Still, because  the two flat bands
As the two sets of quadratic band points  are separated in energy and related by particle-hole symmetry, their low-energy dynamics near $\Gamma$ decouple to leading order. Again, the anomalous Landau level spectrum can be formally derived following appendices \ref{app:QBC_eff},\ref{app:eff} and \ref{app:QBC_LL}. 
The Landau energy levels at the positive-energy quadratic band crossing point are given by
\begin{equation}
E_{n,\pm}^{[4,4](+)}
=
E_{\rm FB}^{(+)}+\frac{4\zeta}{\ell_B^2}\left(n-\frac12\right)
\pm
\frac{4\zeta}{\ell_B^2}\sqrt{n^2-n+1},
\label{eq:Enpm_exact_final0}
\end{equation}
where $\zeta=\frac{243a_{CC}^2}{16}\eta \tilde{t}$ and $n\geq2$. 
The negative energy sector can be obtained from the particle-hole property $E_{n,\pm}^{(-)}=-E_{n,\pm}^{(+)}$ and the corresponding eigenvectors are related to the positive-energy ones by chiral symmetry. Eq.~\eqref{eq:Enpm_exact_final0} exhibits the characteristic Landau quantization of a singular flat band with a quadratic band crossing point. The exact solution combines the cyclotron scale set by the quadratic dispersive band with an additional scale associated with the wave function geometry that controls the anomalous scaling of $E^{[4,4](\pm)}_{n,-}$. The spreading of the flat band Landau levels is given by 
\begin{equation}
    \left|E_{2,-}^{[4,4](\pm)}-E_{\rm FB}^{(\pm)}\right|=\frac{2\zeta}{\ell_B^2}(2\sqrt{3}-3).
\end{equation}
This result is consistent with the spreading $\approx 16\mu$eV obtained numerically in the $[4,4]$triangulene ribbon (see Fig.~\ref{fig:6}f). In the large-$n$ limit,  $R_n=\sqrt{n^2-n+1}\simeq n-\frac{1}{2}+\frac{3}{8n}+\mathcal{O}(n^{-2})$ and one obtains
\begin{equation}
\begin{aligned}
&E_{n,+}^{[4,4](+)}
\approx
E_{\rm FB}^{(+)}+\frac{8\zeta}{\ell_B^2}\left(n-\frac12\right)
,\qquad n\gg1,\\
&E_{n,-}^{[4,4](+)}
\approx
E_{\rm FB}^{(+)}-\frac{3\zeta}{2\ell_B^2}\frac{1}{n},\qquad n\gg1.
\label{eq:large_n_asymptotic}
\end{aligned}
\end{equation}
Specifically, the first sequence scales linearly with $n$ to higher energies away from the flat band whereas the second sequence remains confined near the flat band energy, with a negative shift that scales inversely with $n$. 

Unlike the $[3,3]$ case, where the anomalous LL were located above $E_{FB}$ and below $-E_{FB}$, in the $[4,4]$ case both positive- and negative-energy anomalous Landau states penetrate the central insulating gap. We note that recent theory  work has shown anomalous Landau levels extending into an insulating gap can produce oscillatory magnetization and conductivity \cite{Panda2022,Shao2024} even in the absence of a zero-field Fermi surface, suggesting that [4,4]triangulenes may likewise exhibit quantum oscillatory signatures.

The Landau spectrum at $\Gamma$ scales linearly with the magnetic field since 
$E_{n,\pm}^{(\pm)}-E_{\mathrm{FB}}^{(\pm)}
\propto \ell_B^{-2}\propto |B|$, as dictated by the quadratic structure of the effective Hamiltonian. Therefore, the low-energy magnetic response is controlled by a finite effective mass scale, in contrast to the massless electron motion of the Dirac-Landau quantization.

There are two additional $n=0,1$ levels that do not follow the scaling behavior represented in Eq.~\eqref{eq:Enpm_exact_final0} [see appendix~\ref{app:QBC_LL} in Eq.~\eqref{eq:01LL44}] and whose energy is determined only from the scale set by the dispersive band.

%\orangemark{I can make this section shorter if needed by taking out eq. \ref{eq:Enpm_exact_final0} and \ref{eq:Phi_final_state_ket}}  I would take eq. 45 out

\section{Summary  and Conclusions \label{sec_final}}

The main results of this work can be summarized as follows:
\begin{enumerate}

\item[(i)] Triangulene crystals exhibit distinct Landau spectra
associated with their multiorbital electronic structure.
The $[2,2]$, $[2,3]$, $[3,3]$, and $[4,4]$ crystals connect,
respectively, to graphene, the Lieb lattice, the $p_x$--$p_y$
honeycomb model, and coupled kagome-like sectors.

\item[(ii)] The dispersive Dirac bands develop Landau levels with
square-root magnetic field scaling, summarized by
\begin{equation}
E_{n,\pm}^{[n_a,n_b]}
=
\pm \bar{\epsilon}_0^{[n_a,n_b]}\frac{a_{\mathrm{CC}}}{\ell_B} \sqrt{\lambda n+1},
\label{eq:dirac_landau_spectrum}
\end{equation}
Here, $n=0,1,\ldots$, $\lambda=1$ for the centrosymmetric
crystals and $\lambda=2$ for $[2,3]$. The model-dependent
coefficients $\bar{\epsilon}_0^{[n_a,n_b]}$ are given by the entries in Table~\ref{tabm}. This expression describes the finite-energy sectors while zero-energy states are treated
separately.

\item[(iii)] In the $[2,3]$ crystal, the zero-energy flat band manifold persists in a perpendicular magnetic field. We derive analytical expressions for the finite-energy
Landau spinors and zero-energy states, establishing their correspondence with those of the Lieb lattice.

\item[(iv)] The Landau eigenstates possess characteristic sublattice and orbital structures. In $[3,3]$, the low-energy Dirac states show a valley dependent sublattice--orbital locking, and valley inversion and magnetic field reversal act differently on their internal components. Coupling to the flat band sector produces deviations from the ideal Dirac spectrum at higher Landau indices.

\item[(v)] At the quadratic band crossings of both $[3,3]$ and $[4,4]$, conventional Landau levels originating from the quadratic
dispersive bands coexist with anomalous levels associated with singular flat bands. We obtain both sequences analytically from the effective quadratic Hamiltonians.
Their energy shifts scale linearly with magnetic field while their large-index behavior differs: the conventional
sequence grows as $n$, whereas the anomalous sequence approaches the flat band energy as $1/n$.

\item[(vi)] The analytical spectra connect the spreading of the anomalous Landau levels to the quantum geometry of the flat band states. The effective quadratic Hamiltonians also explain contrasting directions of their shifts in $[3,3]$ and $[4,4]$.

\item[(vii)] In $[4,4]$, the anomalous Landau levels penetrate the zero-field insulating gap in both the valence and conduction flat bands. Thus, magnetic quantization generates in-gap states even though the zero-field
crystal is insulating.

\end{enumerate}

These results establish triangulene crystals as a molecular platform in which Dirac and singular flat band Landau quantization coexist, with their characteristic energy
scales and level sequences controlled by molecular architecture. The connections to the Lieb, $p_x$--$p_y$ honeycomb, and kagome lattice models extend the relevance of our analytical and numerical results beyond
triangulene-based materials, including artificial electronic lattices and polaritonic arrays \cite{Jacqmin2014,jamadi20} with suitably
engineered synthetic gauge fields. More broadly, this work illustrates how magnetic quantization probes both the dispersion and the internal wave function geometry of
multiorbital bands.

\begin{table}[h!]
\centering

\begin{tabularx}{\linewidth}{|>{\centering\arraybackslash}X|>{\centering\arraybackslash}X|>{\centering\arraybackslash}X|>{\centering\arraybackslash}X|>{\centering\arraybackslash}X|}
\hline
\rule{0pt}{3ex}\textbf{} & \textbf{HSP} & $\bar{\epsilon}_0$&\textbf{band}  &\textbf{LL} \\ [0.2ex]
\hline
\rule{0pt}{3ex}{\centering $[2,2]$}& $K$ &$\frac{5}{\sqrt{2}}t_3$ & DB & $\sqrt{n+1}$ \\ [1.2ex]
\hline
\rule{0pt}{3.5ex}{\centering $[2,3]$}& $\Gamma$ & $6\sqrt{\frac{3}{11}}t_3$ & DB & $\sqrt{2n+1}$ \\ [1.2ex]
\hline
\rule{0pt}{3ex}\multirow{3}{*}{\centering $[3,3]$} & $K$ & $\frac{42}{11\sqrt{2}}t_3$ & DB  & $\sqrt{n+1}$ \\ [1ex]
\cline{2-5}
                        \rule{0pt}{2.2ex} & \multirow{2}{*}{\centering $\Gamma$} & \multirow{2}{*}{\centering $\frac{147}{22}\frac{a_{cc}}{\ell_B}t_3$} & DB & $n$ \\ 
\cline{4-5}
                        \rule{0pt}{2.4ex}& & & FB & $1/n$ \\
\hline
\rule{0pt}{3ex}\multirow{3}{*}{\centering $[4,4]$} & $K$ & $\frac{27}{\sqrt{2}}\eta \tilde{t}$ & DB & $\sqrt{n+1}$ \\ [1ex]
\cline{2-5}
                        \rule{0pt}{2.2ex}& \multirow{2}{*}{\centering $\Gamma$} &\multirow{2}{*}{\centering $\frac{81\sqrt{3}}{8}\frac{a_{cc}}{\ell_B}\eta \tilde{t}$} & DB & $n$ \\
\cline{4-5}
                        \rule{0pt}{2.4ex}& & & FB & $1/n$ \\
\hline
\end{tabularx}
\caption{Characteristic scaling behavior and scale prefactor ($\bar{\epsilon}_0=\epsilon_0\,\ell_B/a_{CC}$) of the Landau spectrum for $C_3$ high-fold fermions in honeycomb lattices retrieved from $k\cdot p$ approximation at the two main high-symmetry points (HSP). The hopping terms are $t_3=0.1t$, $\tilde{t}=0.035t$ and $\eta = 0.097$.
For the quadratic band crossings, $\bar{\epsilon}_0$ denotes
the characteristic scale extracted from the coefficients of
$a^2$ and $(a^\dagger)^2$ in the effective Hamiltonian
rather than the full prefactor of either Landau level sequence.
%Note that $\bar{\epsilon}_0$ at the quadratic band crossings was estimated from the $a^2,(a^\dagger)^2$ coefficients after second order expansion of the Hamiltonian.}
}
\label{tabm}
\end{table}

\newpage

{\em Acknowledgments}  We acknowledge João Henriques for fruitful discussions and Nuno Peres for his valuable contribution on the analytical solution for the Landau spectrum of the $[2,3]$ crystal. 
L.M. and J.F.-R. acknowledge support from the European Union (Grant FUNLAYERS-101079184).
L.M acknowledges financial support from the project QUEST2D (PTDC/FIS-MAC/2045/2021). 
J.F.-R. acknowledges financial support from Generalitat Valenciana (Prometeo2021/017 and MFA/2022/045), from SNF Sinergia (Grant Pimag, CRSII5-205987) and MICIN-Spain (Grants  PID2022-141712NB-C22 and PRTR-C1y.I1). 
L.M. developed work within the scope of Portuguese Institute for Nanostructures, Nanomodelling and Nanofabrication (i3N) Projects No. UIDB/50025/2020, No. UIDP/50025/2020, and No. LA/P/0037/2020, financed by national funds through the Fundação para a Ciência e Tecnologia (FCT) and the Ministério da Educação e Ciência (MEC) of Portugal.
We acknowledge the use of OpenAI's ChatGPT as a research aid for theoretical discussions, numerical and coding tasks, and manuscript editing. We have carefully revised all the results. Accordingly, we take full responsibility for the presented material.

\appendix

%%%%%%%%%%%%%%%%%%%

\section{Analytical construction of the magnetic field dependent zero modes of single triangulenes}
\label{SMA}

The construction of the magnetic field dependent reduced ribbon Hamiltonian relies on the zero modes of the individual triangulene monomers. Within the first-neighbor tight-binding description, a $[n]$triangulene hosts $(n-1)$ zero-energy states polarized on the majority sublattice. When a perpendicular magnetic field is introduced, the Peierls phases modify the relative amplitudes and phases of the atomic components of these states. Since the reduced hopping parameters are obtained from the zero mode components at the binding sites, the corresponding molecular states must be determined at finite magnetic field.

In the Landau gauge used throughout this work,
$\mathbf{A}=-By\hat{\mathbf{x}}$, the Peierls phases depend on the vertical position of the monomer inside the ribbon unit cell. We consider an isolated triangulene translated by $n_yR_y$ along the $y$ direction. The position of atom $i$ in the ribbon is
\begin{equation}
    \mathbf{r}_{i}^{(n_y)}
    =
    \left(x_i,y_i+n_yR_y\right),
\end{equation}
where $(x_i,y_i)$ denotes its position with respect to a reference monomer. The Peierls phase associated with a hopping between atoms $i$ and $j$ is
\begin{equation}
\begin{aligned}
    \phi_{ij|_{n_y}}
    &=
    -\frac{eB}{2\hbar}
    \left(y_j+y_i+2n_yR_y\right)
    (x_j-x_i)
    \\
    &=
    \phi_{ij|_{0}}
    -
    \frac{eB}{\hbar}
    n_yR_y(x_j-x_i).
\end{aligned}
\label{eq:phi_monomer}
\end{equation}
Thus, both the magnetic field and the position $n_y$ of the monomer enter the phases of the atomic hopping amplitudes.

To obtain the field dependent zero modes analytically, we retain the leading first neighbor hopping term. The tight-binding Hamiltonian of a triangulene at position $n_y$ reads
\begin{equation}
    H_{\mathrm{TB}}^{(n_y)}(B)
    =
    -t_1
    \sum_{\langle i,j\rangle}
    e^{i\phi_{ij|_{n_y}}}
    c_i^\dagger c_j
    +\mathrm{h.c.},
\label{eq:Htri_B}
\end{equation}
and remains bipartite, since the magnetic field only introduces complex phases in the hopping amplitudes.

We first consider a type-$A$ $[n_a]$triangulene, for which $A$ is the majority sublattice, hosting  The $n_a-1$ zero modes have vanishing amplitude on the minority $B$ sublattice and can be written as
\begin{equation}
    \ket{\widetilde{\psi}_{a}}
    =
    \frac{1}{\mathcal N_{a}^{(n_y)}}
    \sum_{i\in A}
    \alpha_{i,a}^{(n_y)}
    \ket{i},
    \quad
    a=1,\ldots,n_a-1,
\label{eq:ZM_expansion}
\end{equation}
where $\mathcal N_{a}^{(n_y)}$ is the normalization constant and
\begin{equation}
    \langle j|
    \widetilde{\psi}_{a}
    \rangle=0,
    \qquad
    j\in B.
\end{equation}
The zero-energy Schr\"odinger equation is
\begin{equation}
    H_{\mathrm{TB}}^{(n_y)}(B)
    \ket{\widetilde{\psi}_{a}}
    =0.
\label{eq:ZM_Schrodinger}
\end{equation}
The nontrivial conditions are obtained by projecting Eq.~\eqref{eq:ZM_Schrodinger} onto the minority sublattice sites. For each $j\in B$,
\begin{equation}
    \sum_{i\in\mathcal{N}(j)}
    e^{i\phi_{ji|_{n_y}}}
    \alpha_{i,a}^{(n_y)}
    =0,
    \qquad
    j\in B,
\label{eq:ZM_local}
\end{equation}
where $\mathcal{N}(j)$ denotes the set of first-neighbor majority sublattice sites connected to $j$. Eq.~\eqref{eq:ZM_local} is the relation used to determine the magnetic field dependent components of each zero mode.

The number of majority- and minority-sublattice sites satisfies $N_A-N_B=n_a-1$ so that the local conditions in Eq.~\eqref{eq:ZM_local} define an $(n_a-1)$-dimensional zero-energy subspace. We fix one coefficient for each solution,
\begin{equation}
    \alpha_{i_0,a}^{(n_y)}=1,
\label{eq:ZM_reference}
\end{equation}
and use the normalization condition for the resulting state.

When $n_a>2$, the local conditions determine a degenerate $(n_a-1)$-dimensional zero-energy subspace. After fixing the overall scale through Eq.~\eqref{eq:ZM_reference}, $(n_a-2)$ independent conditions are required to select the states forming the zero-mode basis. In the analytical construction used here, these conditions are obtained from the magnetic mirror symmetries of the isolated triangulene. For pairs of sites $p_s$ and $q_s$ related by the corresponding spatial mirrors, the perpendicular magnetic field requires the mirror operation to be combined with complex conjugation in the spinless problem. The phases of the corresponding amplitudes may differ, while their probability weights remain equal,
\begin{equation}
    \left|
    \alpha_{p_s,a}^{(n_y)}
    \right|
    =
    \left|
    \alpha_{q_s,a}^{(n_y)}
    \right|,
    \qquad
    s=1,\ldots,n_a-2.
\label{eq:ZM_mirror_general}
\end{equation}
Together with Eqs.~\eqref{eq:ZM_local} and \eqref{eq:ZM_reference}, these conditions select the independent coefficient sets
$\alpha_{i,a}^{(n_y)}$ entering Eq.~\eqref{eq:ZM_expansion}.

The same construction applies to a type-$B$ triangulene, with the zero modes polarized on the $B$ sublattice. Thus, for a general $[n_a,n_b]$triangulene dimer we obtain
\begin{equation}
    \left\{
    \ket{\widetilde{\psi}_{a}}
    \right\}_{a=1}^{n_a-1},
    \qquad
    \left\{
    \ket{\widetilde{\psi}_{b}}
    \right\}_{b=1}^{n_b-1}.
\label{eq:ZM_sets}
\end{equation}

The components of these states at the binding sites are then used to construct the magnetic field dependent reduced hopping amplitudes. For a hopping process connecting a type-$A$ and a type-$B$ triangulene, the general matrix element reads
\begin{equation}
\tilde{t\,}^{ab}_{m,n_y}=\sum_{i,j\big|_{m,n_y}}\tilde{t}_3\tilde{\psi}_a(i)\tilde{\psi}_b^*(j)
\label{eq:appmagt}
\end{equation}
where tilde denotes the magnetic field dependent quantities. Both the Peierls phases associated with hopping between neighboring triangulenes and through the field dependence of the zero mode wave functions enter 
the magnetic field reduced ribbon Hamiltonian.

\section{Valley transformation and ladder-operator structure \label{sec:valley-vs-a}}
In this appendix we show the relation between the $k\cdot p$ Hamiltonians at the $K$ and $K'$ valleys.
We start from the time-reversal property of spinless tight-binding Hamiltonians with real hopping amplitudes,  such as the zero-field Bloch Hamiltonian of the 2D triangulene crystals [of Eq.~\eqref{H2Def}]
\begin{equation}
H_{\mathrm{r}}(\mathbf{k})
=
H_{\mathrm{r}}^{*}(-\mathbf{k}),
\end{equation}
where the subscript ($\mathrm{r}$) refers to the Hamiltonian written in a real orbital basis. In this representation, the time-reversal operator is a simple complex conjugation, $\Theta_{\mathrm{r}}=\mathcal{K}$. 
In the effective triangulene basis used in this work, the low-energy orbitals carry $C_3$ character. These two orbitals are exchanged by complex conjugation. Therefore, in the orbital basis, time reversal is represented as $\Theta = U_{\mathcal T} K $, where $U_{\mathcal T}$ is the orbital swap. Explicitly in the $[3,3]$triangulene minimal model, 
\begin{equation}
U_{\mathcal T}=
\begin{pmatrix}
0&1&0&0\\
1&0&0&0\\
0&0&0&1\\
0&0&1&0
\end{pmatrix}.
\end{equation}
in the $(a_+,a_-,b_+,b_-)$ orbital basis.
Thus, the time-reversal symmetry condition becomes
\begin{equation}
H(\mathbf{k}) =
U_{\mathcal T} H^*(-\mathbf{k}) U_{\mathcal T}^{\dagger}.
\end{equation}

We define
\begin{equation}
H_K(\mathbf{q})= H(\mathbf{K}+\mathbf{q}), 
\qquad
H_{K'}(\mathbf{q})= H(\mathbf{K'}+\mathbf{q}) .
\end{equation}
Since $K'$ is related to $K$ by inversion in momentum space, $\mathbf K'=-\mathbf K$,
the expansion around $K'$ can be expressed as
\begin{equation}
H_{K'}(\mathbf{q})
=
U_{\mathcal T} H_K^*(-\mathbf{q}) U_{\mathcal T}^{\dagger},
\end{equation}
where $\mathbf{q}$ is the momentum relative to the valley.

We now analyze the consequences of this relation for the $k\cdot p$ expansion. Introducing
\begin{equation}
q_\pm = q_x \pm i q_y,
\end{equation}
the transformation $\mathbf{q}\rightarrow -\mathbf{q}$ together with complex conjugation gives
\begin{equation}
q_+ \rightarrow -q_-,
\qquad
q_- \rightarrow -q_+ .
\end{equation}
Thus, up to linear order in $\mathbf q$, the Hamiltonian at $K'$ is obtained from that at $K$ by exchanging $q_+$ and $q_-$, with an additional minus sign. The full transformation includes the unitary matrix $U_{\mathcal T}$, which exchanges the internal $+$ and $-$ orbital labels.

We now connect this result to the Landau-level representation introduced in Section~\ref{SecII.C}. Using the dimensionless operators
\begin{equation}
Q = \frac{y}{\ell_B} + q_x \ell_B,
\qquad
P = \frac{\ell_B}{\hbar} p_y,
\end{equation}
and the definition for the ladder operators, we obtain 
\begin{equation}
q_x \rightarrow \frac{Q}{\ell_B},
\qquad
q_y \rightarrow \frac{P}{\ell_B}.
\end{equation}
Consequently,
\begin{equation}
q_+ \rightarrow \frac{\sqrt{2}}{\ell_B}a,
\qquad
q_- \rightarrow \frac{\sqrt{2}}{\ell_B}a^\dagger .
\end{equation}
Combining this with the valley transformation above, we find the valley transformation maps the ladder operators as
\begin{equation}
q_+ \rightarrow -q_-
\quad \Longrightarrow \quad
a \rightarrow -a^\dagger.
\end{equation}
Therefore, the full relation between the Hamiltonian at the valleys yields
\begin{equation}
H_{K'}^B(a,a^\dagger)
=
U_{\mathcal T}
H_K^B(-a^\dagger,-a)
U_{\mathcal T}^{\dagger},
\end{equation}
In the two-band Dirac Hamiltonian, the  phase can be absorbed in a basis redefinition and valley transformation exchanges $
a \leftrightarrow a^\dagger $. 
In a multiorbital Hamiltonian, the exact transformation is
\begin{equation}
a \leftrightarrow -a^\dagger,
\qquad
+\leftrightarrow - .
\end{equation}

\section{Analytical derivation of the Landau states for the Lieb lattice}\label{SMB}

We analyze the Landau eigenstates of the Lieb lattice starting from the low-energy Hamiltonian approximation at the $M$-point
\begin{equation}
   H_k = i\begin{pmatrix}
0 & k_x & k_y\\
-k_x & 0 & 0\\
-k_y & 0 & 0
\end{pmatrix}
\end{equation}
We write this Hamiltonian in terms of the harmonic-oscillator ladder operators $a,a^\dagger$ acting on the Landau-level Hilbert space $\mathcal H=\operatorname{span}\{\ket{n}\}_{n\ge0}$. Working in the three-component space $\mathcal H^{(3)}=\mathcal H\oplus\mathcal H\oplus\mathcal H$ with spinor in the usual convention $\Phi=(\ket{\Phi_n^{A}},\ket{\Phi_n^{B}},\ket{\Phi_n^{C}})^{\mathsf T}$, the magnetic Hamiltonian reads 
\begin{equation}
   H = \frac{\epsilon_0^{\text{Lieb}}}{\sqrt{2}}\begin{pmatrix}
0 & -i(a+a^\dagger) & a-a^\dagger\\[4pt]
i(a+a^\dagger) & 0 & 0\\
a^\dagger-a & 0 & 0
\end{pmatrix},
\end{equation}
with $\epsilon_0^{\text{Lieb}}=\frac{a_{\rm L}t}{\,\ell_B}$ and $a_{\rm L}$ the Lieb lattice constant. The eigenproblem is the system of equations
\begin{equation}
\begin{aligned}
\frac{\epsilon_0^{\text{Lieb}}}{\sqrt{2}}\left(-i(a+a^\dagger)\ket{\Phi_n^{B}} + (a-a^\dagger)\ket{\Phi_n^{C}}\right)& = E_n\ket{\Phi_n^{A}}, \\
\frac{i\epsilon_0^{\text{Lieb}}}{\sqrt{2}}(a+a^\dagger)\ket{\Phi_n^{A}}& = E_n\ket{\Phi_n^{B}}, \\
\frac{\epsilon_0^{\text{Lieb}}}{\sqrt{2}}(a^\dagger-a)\ket{\Phi_n^{A}}& = E_n\ket{\Phi_n^{C}}. 
\end{aligned}
\end{equation}
We may write the equation for A-orbital as 
\begin{equation}
\frac{\left(\epsilon_0^{\text{Lieb}}\right)^2}{2}\left(|a+a^\dagger|^2+|a-a^\dagger|^2\right)\ket{\Phi_n^{A}}
=E_n^2\,\ket{\Phi_n^{A}}.
\end{equation}
The ladder operator term can be further simplified to 
$|a+a^\dagger|^2+|a-a^\dagger|^2=2\big(aa^\dagger+a^\dagger a\big)=4n+2$
where $n=a^\dagger a$. Therefore the Landau level scaling of the Lieb lattice follows
\begin{equation}
   E_{n}^\pm=\pm\epsilon_0^{\text{Lieb}}\sqrt{\,2n+1\,}\;,\qquad n=0,1,2,\dots\;
\end{equation}
and the closed-form of the $A$-component as the number eigenstate $\psi_A=\ket{n}$. The remaining two components can be retrieved from the last two equations of the eigensystem so that the normalized eigenvectors read

\begin{equation}
\ket{\Phi_{n}^\pm} \;=\; \frac{1}{\sqrt{4n+2}}
\begin{pmatrix}
\ket{n} \\[6pt]
\displaystyle \pm i\sqrt{n}\ket{n-1} \pm i\sqrt{n+1}\ket{n+1} \\[6pt]
\displaystyle \mp\sqrt{n+1}\ket{n+1} \;\pm\; \sqrt{n}\ket{n-1}
\end{pmatrix}
\end{equation}
The general finite-energy construction above remains valid at $n=0$ even though $\ket{n-1}$ is ill-defined. If $\ket{\Phi_n^{A}}=\ket{0}$ then we explicitly have
$\ket{\Phi_{0}^\pm}=(\ket{0},\pm i\ket{1},\mp\ket{1})^\mathsf{T}/\sqrt{3}$. Looking for further solutions of the form $\ket{\Phi_n^{B(C)}}=\ket{0}$ does not satisfy the last two equations of the eigenproblem. 

Since the bands of the Lieb lattice include a triple band crossing at the M-point with a flat band, a magnetic manifold at $E=0$ is also expected.   
The last two equations of the eigensystem yield
\begin{equation}
i\epsilon_0^{\text{Lieb}}(a+a^\dagger)\ket{\mathcal{Z}_{0}^A}=0,\qquad \epsilon_0^{\text{Lieb}}(a^\dagger-a)\ket{\mathcal{Z}_{0}^A}=0.
\end{equation}
These two operator equalities imply  $a\ket{\mathcal{Z}_{0}^A}=0$ and $a^\dagger\psi_A=0$, whose only solution in $\mathcal H$ is the trivial vector $\ket{\mathcal{Z}_{0}^A}=0$. Thus, every zero-energy state must have vanishing weight on the A-sublattice. The remaining equation gives the operator relation coupling the B,C components 
\begin{equation}\label{zerocLieb1}
(a-a^\dagger)\ket{\mathcal{Z}_{0}^C} \;=\; i(a+a^\dagger)\ket{\mathcal{Z}_{0}^B}.
\end{equation}
and does not fix a single Landau index $n$. Consequently the kernel of $H$ is infinite dimensional and the zero modes are generically superpositions of many $\ket{n}$ levels on these two sublattices. 
To express the constraint in Eq.~\eqref{zerocLieb1},
we introduce the combinations
\begin{equation}
\ket{\mathcal{Z}_n^{BC,+}}
=\frac{\ket{\mathcal{Z}_n^C}
+i\ket{\mathcal{Z}_n^B}}{\sqrt{2}},
\quad
\ket{\mathcal{Z}_n^{BC,-}}
=\frac{\ket{\mathcal{Z}_n^C}
-i\ket{\mathcal{Z}_n^B}}{\sqrt{2}}.
\end{equation}
The labels $+$ and $-$ distinguish these combinations.
Equation~\eqref{zerocLieb1} then becomes
\begin{equation}\label{zerocLieb2}
a\ket{\mathcal{Z}_n^{BC,-}}
=a^\dagger\ket{\mathcal{Z}_n^{BC,+}}.
\end{equation}
For $n\geq1$, a convenient family of solutions is
\begin{equation}
\ket{\mathcal{Z}_n^{BC,+}}=z_1\ket{n-1},
\quad
\ket{\mathcal{Z}_n^{BC,-}}=z_2\ket{n+1}.
\end{equation}
The ladder-operator constraint requires
\begin{equation}
z_1=\sqrt{\frac{n+1}{n}}\,z_2.
\end{equation}
Since the transformation between the original components
and these combinations is unitary, normalization requires
$|z_1|^2+|z_2|^2=1$. Choosing the overall phase so that
both coefficients are real and positive gives
\begin{align}
\ket{\mathcal{Z}_n^{BC,+}}
&=\sqrt{\frac{n+1}{2n+1}}\ket{n-1},
\\
\ket{\mathcal{Z}_n^{BC,-}}
&=\sqrt{\frac{n}{2n+1}}\ket{n+1},
\end{align}
and the general solution for the zero-energy states in the original basis follow
\begin{equation}
 \ket{\mathcal{Z}_n} \;=\; \frac{1}{\sqrt{4n+2}}
\begin{pmatrix}
0\\[6pt]
\mp i\sqrt{n}\ket{n+1} \pm i\sqrt{n+1}\ket{n-1}\\
\pm\sqrt{n}\ket{n+1} \pm \sqrt{n+1}\ket{n-1}
\end{pmatrix},
\end{equation}
for $n>0$. The $n=0$ solution is not included in the general construction since the $\ket{n-1}$ term is not well defined.  Setting either $\ket{\mathcal{Z}_n^{B(C)}} \equiv \ket{0}$ breaks the relation in Eq.~(\ref{zerocLieb1}). In fact, the vaccuum state should be applied to the combination basis so that Eq.~(\ref{zerocLieb2}) is satisfied. The solution when $\ket{\mathcal{Z}_n^{BC}} \equiv \ket{0}$  is already included in the general construction for $n=1$. We assume solutions of the form $\ket{\mathcal{Z}_n^{BC}}^* \equiv \ket{0}$ so that $\ket{\mathcal{Z}_n^{BC}}^* =0$ and the missing normalized eigenvector is 
$\ket{\mathcal{Z}_{0}}=(0,i\ket{0},\ket{0})^\mathsf{T}/\sqrt{2}$. Boundary effects associated with $a\ket{0}=0$ produce the special $n=0$ states in both sectors: a finite-energy doublet $\ket{\Phi_{0}^\pm}$ and the zero-mode $\ket{\mathcal{Z}_{0}}$.

\section{Effective Hamiltonians from low-energy projection}
\label{app:eff}
We now describe a procedure to derive effective $2\times2$ $k\cdot p$ Hamiltonians, that will become useful to analyze the Landau Levels for the $[3,3]$ and $[4,4]$ cases close to the Dirac points at $K$ and $K'$ and for the quadratic band
crossing points at $\Gamma$.
These  effective Hamiltonians  are obtained by projecting the full Bloch Hamiltonian onto the
low-energy subspace of interest. 

We first diagonalize
$H^{(0)}=H(\mathbf{k}_0)$  in Eq.~\eqref{eq:kp} and identify the degenerate manifold associated with the band crossing of interest. Denoting by $P$ the projector onto this subspace and by $Q=1-P$ its complement, the effective Hamiltonian acting within the $P$ subspace can be constructed perturbatively by eliminating
the remote states.
Up to second order in momentum, L\"owdin perturbation theory \cite{Lowdin1951} gives
\begin{equation}
\begin{split}
H_{\mathrm{eff}}
={}& E_0 P + P H^{(1)} P + P H^{(2)} P \\
&- P H^{(1)} Q
\left(QH^{(0)}Q-E_0\right)^{-1} 
Q H^{(1)} P
+ \mathcal{O}(q^3).
\end{split}
\label{eq:lowdin_general}
\end{equation}
where $E_0$ is the energy of the manifold at $\mathbf{k}_0$. The last term accounts for virtual transitions to the remote bands. Once expressed in a basis spanning the $P$ subspace, Eq.~\eqref{eq:lowdin_general} yields the reduced $k\cdot p$
Hamiltonian.

The order at which the first nonvanishing term appears depends on the symmetry and nature of the band crossing. For the $[3,3]$ crystal at the $K$ and $K'$ points, the subspace is built from the zero-energy Dirac manifold. In this case $PH^{(1)}P\neq0$, and the leading projected Hamiltonian is linear in momentum,
\begin{equation}
H_{\mathrm{eff}}^{K(K')}= P H^{(1)} P+\mathcal{O}(q^2),
\end{equation}
giving the two-component Dirac Hamiltonian. The remaining states form a remote high-energy sector whose coupling to the Dirac manifold generates higher-order corrections.

At the $\Gamma$ point of both $[3,3]$ and $[4,4]$ crystals, the relevant two-dimensional manifold is formed by the flat band and the dispersive band that are degenerate at the quadratic band crossing point. In these cases, symmetry enforces
\begin{equation}
P H^{(1)} P=0,
\end{equation}
so that the leading effective Hamiltonian is quadratic,
\begin{equation}
\begin{split}
H_{\mathrm{eff}}^{\Gamma}
=\,&
E_{\mathrm{FB}}P
+
P H^{(2)}P\\
&-P H^{(1)}Q
\left(
QH^{(0)}Q-E_{\mathrm{FB}}
\right)^{-1}
QH^{(1)}P
+
\mathcal{O}(q^3).
\end{split}
\label{eq:lowdin_gamma}
\end{equation}
The second term in Eq.~\eqref{eq:lowdin_gamma} represents the direct quadratic contribution within the crossing subspace,
whereas the third arises from virtual transitions through the
remaining bands. Both contributions must in general be retained.

After choosing an orthonormal basis for the projected manifold, $H_{\mathrm{eff}}$ becomes a $2\times2$
Hamiltonian. The perpendicular magnetic field is then introduced through minimal coupling, $\hbar\,\mathbf{q}\rightarrow\boldsymbol{\Pi}$, followed by the representation of $\pi_x$ and $\pi_y$ in terms of ladder operators. This leads to $(a,a^\dagger)$ couplings in the Dirac valleys with $\Delta n=\pm1$, $(a,a^\dagger)^2$ couplings in the quadratic band crossings with $\Delta n=\pm2$.

\section{Dirac and flat band sectors of the $[3,3]$triangulene model at $K$}
\label{app:33K}

Here we analyze the coupling between the Dirac Landau levels
and the remote flat band subspace of the $[3,3]$ crystal. The
general projection procedure used to construct reduced
$k\cdot p$ Hamiltonians is discussed in Appendix~\ref{app:eff}.

We expand the full Hamiltonian to first order around
$\mathbf K$, with $\mathbf k=\mathbf K+\mathbf q$ and use
the eigenbasis of $H(\mathbf K)$. The two zero-energy states
span the subspace
\begin{equation}
\mathcal H_G=\mathrm{span}\{|a_-\rangle,|b_+\rangle\},
\end{equation}
and the remaining two states form the remote sector
$\mathcal H_F$. In this basis, the Hamiltonian takes the block
form
\begin{equation}
H_K(\mathbf q)=
\begin{pmatrix}
H_D & V\\
V^\dagger & H_{FB}
\end{pmatrix},
\label{eq:K_block}
\end{equation}
with
\begin{equation}
H_D=-\sqrt{3}\epsilon_0 a_{CC}
\begin{pmatrix}
0&q_+\\
q_-&0
\end{pmatrix},
\quad q_\pm=q_x\pm iq_y ,
\label{eq:HG}
\end{equation}
\begin{equation}
H_{FB}=
\begin{pmatrix}
E_{\rm FB}^{(+)}&2i{\cal E}_0\sqrt{3}a_{CC} q_y\\
-2i{\cal E}_0\sqrt{3}a_{CC} q_y&E_{\rm FB}^{(-)}
\end{pmatrix}.
\label{eq:HF}
\end{equation}
and ${\cal E}_0=(7\sqrt{3}/2)t_{\rm eff}^{[3,3]}$.
Matrix $V$ is linear in momentum and couples the
Dirac and remote sectors.
For $B>0$, using
$q_+\rightarrow\sqrt{2}a/\ell_B$ and
$q_-\rightarrow\sqrt{2}a^\dagger/\ell_B$, the coupling
becomes
\begin{equation}
V(B)=
\frac{{\cal E}_0\sqrt{3}a_{CC}}{\ell_B}
\begin{pmatrix}
a^\dagger&a^\dagger\\
a&-a
\end{pmatrix}.
\label{eq:V33}
\end{equation}
In the decoupled limit, the Dirac block gives the scaling
\begin{equation}
E_{K,n}^{[3,3]}
=\pm\epsilon_0^{[3,3]}\sqrt{n+1},
\end{equation}
with the corresponding states discussed in the main text.
We estimate the validity of this description from
the ratio between the matrix elements of $V$ and the energy
separation to remote bands. For a Landau level $n$,
\begin{equation}
\lambda_n\sim
\frac{\epsilon_0\sqrt{3}a_{CC}}{E_{\rm FB}\ell_B}\sqrt{n+1}
\sim\frac{|E_{K,n}^{[3,3]}|}{E_{\rm FB}},
\label{eq:mixing33}
\end{equation}
up to a numerical prefactor of magnitude close to one. Hence, the Dirac and flat band LL become asymptotically decoupled for small $n$ and weak magnetic field, while their hybridization
increases as $\sqrt{nB}$. This accounts for the progressive
deviation of the full Landau spectrum with respect to the Dirac-Landau quantization seen in Fig.~\ref{fig:4}.

\section{Effective Hamiltonians at the quadratic band crossing points}
\label{app:QBC_eff}

In this appendix we derive the Hamiltonians that
describe the quadratic band crossing points of the $[3,3]$
and $[4,4]$ triangulene crystals at $\Gamma$. The general projection procedure is described in Secion~\ref{methods}. Here,   we specify the results for each model required for the Landau quantization in Appendix~\ref{app:QBC_LL}. In both cases, we focus on the positive-energy crossing. The negative-energy sector follows from chiral symmetry.

\subsection{$[3,3]$triangulene}

In the zero mode basis $(a_+,a_-,b_+,b_-)$, the reduced
Hamiltonian has the bipartite form
\begin{equation}
H^{[3,3]}(\mathbf k)=
\begin{pmatrix}
0 & \tau(\mathbf k)\\
\tau^\dagger(\mathbf k)&0
\end{pmatrix},
\end{equation}
where
\begin{equation}
\tau(\mathbf k)=t_{\rm eff}^{[3,3]}
\begin{pmatrix}
f(\phi_1,\phi_2)&f(\phi_1+\theta,\phi_2-\theta)\\
f(\phi_1-\theta,\phi_2+\theta)&f(\phi_1,\phi_2)
\end{pmatrix},
\end{equation}
\begin{equation}
f(x,y)=1+e^{ix}+e^{iy},\quad
\theta=\frac{2\pi}{3},\quad
t_{\rm eff}^{[3,3]}=\frac{2t_3}{11}.
\end{equation}
At $\Gamma$, $\tau(0)=3t_{\rm eff}^{[3,3]}I_2$, and the
positive-energy crossing occurs at
\begin{equation}
E_{\rm FB}^{(+)}=3t_{\rm eff}^{[3,3]}=\frac{6t_3}{11}.
\end{equation}
A convenient orthonormal basis of this two-dimensional
subspace is
\begin{equation}
|u_1\rangle=\frac{1}{\sqrt2}(1,0,1,0)^T,\qquad
|u_2\rangle=\frac{1}{\sqrt2}(0,1,0,1)^T.
\end{equation}
Applying the second-order projection discussed in Section~\ref{methods}
and expressing the result after a constant unitary transformation within the
degenerate subspace,
\begin{equation}
H_{\rm eff}^{[3,3](+)}(\mathbf k)
=
E_{\rm FB}^{(+)}I_2
-\gamma
\begin{pmatrix}
k^2&k_+^2\\
k_-^2&k^2
\end{pmatrix},
\label{eq:Heff33}
\end{equation}
with $k_\pm=k_x\pm ik_y$ and
$\gamma=\frac{147}{44}t_3a_{\rm CC}^2$. 
The two eigenvalues are
\begin{equation}
E_{\rm FB}^{(+)},\qquad
E_{\rm FB}^{(+)}-2\gamma k^2,
\end{equation}
which shows explicitly the flat and the dispersive band touching at $\Gamma$.

\subsection{$[4,4]$triangulene}

Likewise, the reduced Hamiltonian of the $[4,4]$ crystal can be written as
\begin{equation}
H^{[4,4]}(\mathbf k)=
\begin{pmatrix}
0&\tau(\mathbf k)\\
\tau^\dagger(\mathbf k)&0
\end{pmatrix},
\quad
\tau(\mathbf k)=\tilde t\sum_{j=0}^{2}M_j e^{i\phi_j},
\end{equation}
where $\phi_j=\mathbf{k}\cdot\mathbf{R}_j$ are the Bloch
phases associated with the lattice translations
with $\mathbf{R}_0=\mathbf{0}$, and the matrices $M_j$ define the reduced $[4,4]$ tight-binding model. At $\Gamma$, the relevant two-dimensional eigenspace of $\tau(0)$ is spanned by
\begin{equation}
\xi_1=\frac{1}{\sqrt2}(1,-1,0)^T,\qquad
\xi_2=\frac{1}{\sqrt6}(1,1,-2)^T,
\end{equation}
with eigenvalue
\begin{equation}
E_{\rm FB}^{(+)}=\tilde t(1-2\eta).
\end{equation}
The corresponding states of the full bipartite Hamiltonian
are
\begin{equation}
|u_i\rangle=\frac{1}{\sqrt2}
\begin{pmatrix}\xi_i\\ \xi_i\end{pmatrix},
\qquad i=1,2.
\end{equation}
Projection to second order in momentum gives
\begin{equation}
H_{\rm eff}^{[4,4](+)}(\mathbf k)
=
E_{\rm FB}^{(+)}I_2+
\zeta
\begin{pmatrix}
h_{11}&h_{12}\\
h_{12}&h_{22}
\end{pmatrix},
\label{eq:Heff44}
\end{equation}
where
\begin{align}
h_{11}&=k_x^2+2\sqrt3 k_xk_y+3k_y^2,\\
h_{12}&=-\left(\sqrt3 k_x^2+2k_xk_y-\sqrt3 k_y^2\right),\\
h_{22}&=3k_x^2-2\sqrt3 k_xk_y+k_y^2,
\end{align}
and $\zeta=\frac{243}{16}\eta\tilde t\,a_{\rm CC}^2$. The quadratic matrix satisfies
\begin{equation}
\det h=0,\qquad {\rm Tr}\,h=4k^2.
\end{equation}
Consequently, the two energy bands close to the positive crossing are 
\begin{equation}
E_{\rm FB}^{(+)},\qquad
E_{\rm FB}^{(+)}+4\zeta k^2.
\end{equation}
Eq.~\eqref{eq:Heff33} and Eq.~\eqref{eq:Heff44} describe the flat-quadratic band crossing with opposite curvature in the dispersive band. As shown in Appendix~\ref{app:QBC_LL}, this difference determines whether the anomalous Landau level sequence develops above or below $E_{\rm FB}^{(+)}$.

\section{Landau quantization of the quadratic band crossing points}
\label{app:QBC_LL}

We now quantize the two effective Hamiltonians derived in Appendix~\ref{app:QBC_eff}. For $B>0$, we use
\begin{equation}
k_+\rightarrow\frac{\sqrt{2}}{\ell_B}a,\quad
k_-\rightarrow\frac{\sqrt{2}}{\ell_B}a^\dagger,\quad
k^2\rightarrow\frac{2N+1}{\ell_B^2}.
\end{equation}
with $N=a^\dagger a$. Since the off-diagonal terms of these Hamiltonians are
quadratic in momentum, they couple oscillator states whose
indices differ by two.

\subsection{$[3,3]$triangulene}

Applying the above substitution to
Eq.~\eqref{eq:Heff33} gives
\begin{equation}
H_{\rm eff}^{[3,3](+)}(B)
=
E_{\rm FB}^{(+)}I_2
-\frac{\gamma}{\ell_B^2}
\begin{pmatrix}
2N+1 & 2a^2\\
2(a^\dagger)^2 & 2N+1
\end{pmatrix}.
\label{eq:H33B}
\end{equation}
For $n\geq2$, the invariant subspace is
\begin{equation}
\mathcal H_n^{[3,3]}
=
\operatorname{span}\{
|u_2\rangle|n\rangle,\,
|u_1\rangle|n-2\rangle
\}.
\end{equation}
Within this subspace, Eq.~\eqref{eq:H33B} reduces to
\begin{equation}
E_{\rm FB}^{(+)}I_2-
\frac{\gamma}{\ell_B^2}
\begin{pmatrix}
2n+1&2\sqrt{n(n-1)}\\
2\sqrt{n(n-1)}&2n-3
\end{pmatrix}.
\end{equation}

The two Landau-level sequences are 
\begin{align}
E_{n,+}^{[3,3](+)}
&=
E_{\rm FB}^{(+)}
-\frac{\gamma}{\ell_B^2}
\left(2n-1+2R_n\right),\\
E_{n,-}^{[3,3](+)}
&=
E_{\rm FB}^{(+)}
-\frac{\gamma}{\ell_B^2}
\left(2n-1-2R_n\right),
\quad n\geq2.
\end{align}
where $R_n=\sqrt{n^2-n+1}$. The first sequence evolves from the quadratic dispersive
band and the second is the anomalous sequence associated with the singular flat band.
For large $n$,
\begin{align}
E_{n,+}^{[3,3](+)}-E_{\rm FB}^{(+)}
&\simeq
-\frac{4\gamma}{\ell_B^2}
\left(n-\frac12\right),\\
E_{n,-}^{[3,3](+)}-E_{\rm FB}^{(+)}
&\simeq
+\frac{3\gamma}{4\ell_B^2}\frac1n.
\label{eq:sfbkp33}
\end{align}
Thus, the two sets display respectively the
conventional $nB$ and the anomalous $B/n$
scaling. In particular, the anomalous levels approach the
positive-energy flat band from above.

The two unpaired $n=0,1$ states follow
\begin{equation}
E_0^{[3,3](+)}
=E_{\rm FB}^{(+)}-\frac{\gamma}{\ell_B^2},
\quad
E_1^{[3,3](+)}
=E_{\rm FB}^{(+)}-\frac{3\gamma}{\ell_B^2}.
\end{equation}

\subsection{$[4,4]$triangulene}

After minimal coupling, Eq.~\eqref{eq:Heff44} can be
written as
\begin{equation}
H_{\rm eff}^{[4,4](+)}(B)
=
D_N I_2
+\frac{\zeta}{\ell_B^2}
\left[a^2M+(a^\dagger)^2M^\dagger\right].
\label{eq:H44B}
\end{equation}
where
\begin{equation}
D_N=E_{\rm FB}^{(+)}
+\frac{4\zeta}{\ell_B^2}
\left(N+\frac12\right),
\quad
M=2e^{i\pi/3}
\begin{pmatrix}
-1&i\\
i&1
\end{pmatrix}.
\end{equation}
It is convenient to introduce
\begin{equation}
|\chi_1\rangle=
\frac{-i|u_1\rangle+|u_2\rangle}{\sqrt2},
\qquad
|\chi_2\rangle=
\frac{i|u_1\rangle+|u_2\rangle}{\sqrt2}.
\end{equation}
These spinors satisfy
\begin{equation}
M|\chi_2\rangle=0,\qquad
M^\dagger|\chi_1\rangle=0,
\end{equation}
together with
\begin{equation}
M|\chi_1\rangle=4e^{i\pi/3}|\chi_2\rangle,\qquad
M^\dagger|\chi_2\rangle=4e^{-i\pi/3}|\chi_1\rangle.
\end{equation}
Consequently, for $n\geq2$ the problem closes within the two-dimensional subspace

\begin{equation}
\mathcal H_n^{[4,4]}
=
\operatorname{span}\{
|\chi_1\rangle|n\rangle,\,
|\chi_2\rangle|n-2\rangle
\}.
\end{equation}
The corresponding $2\times2$ problem has diagonal
elements
\begin{equation}
d_n=E_{\rm FB}^{(+)}
+\frac{4\zeta}{\ell_B^2}
\left(n+\frac12\right)
\end{equation}
and the right off-diagonal element is
\begin{equation}
c_n
=
\frac{4\zeta}{\ell_B^2}
e^{-i\pi/3}\sqrt{n(n-1)}.
\end{equation}
The corresponding eigenvalues are
\begin{equation}
E_{n,\pm}^{[4,4](+)}
=
E_{\rm FB}^{(+)}
+\frac{4\zeta}{\ell_B^2}
\left(n-\frac12\pm R_n\right),
\quad n\geq2.
\end{equation}
For large $n$,
\begin{align}
E_{n,+}^{[4,4](+)}-E_{\rm FB}^{(+)}
&\simeq
\frac{8\zeta}{\ell_B^2}
\left(n-\frac12\right),\\
E_{n,-}^{[4,4](+)}-E_{\rm FB}^{(+)}
&\simeq
-\frac{3\zeta}{2\ell_B^2}\frac1n.
\end{align}
The anomalous sequence approaches the positive-energy flat band from below, in contrast with the $[3,3]$ case.

The two unpaired $n=0,1$ states have energies
\begin{equation}
E_0^{[4,4](+)}
=E_{\rm FB}^{(+)}+\frac{2\zeta}{\ell_B^2},
\quad
E_1^{[4,4](+)}
=E_{\rm FB}^{(+)}+\frac{6\zeta}{\ell_B^2}.
\label{eq:01LL44}
\end{equation}

For both models the negative-energy spectrum follows
directly from chiral symmetry,
\begin{equation}
E_{\alpha,n}^{(-)}=-E_{\alpha,n}^{(+)}.
\end{equation}
The results make transparent the relation between
band curvature and the direction of the anomalous scaling
\begin{align}
E_{\rm QB}^{[3,3]}(\mathbf k)-E_{\rm FB}^{(+)}
&=-2\gamma k^2,
&
E_{\rm FB,n}^{[3,3](+)}-E_{\rm FB}^{(+)}
&\simeq+\frac{3\gamma}{4\ell_B^2n},
\\
E_{\rm QB}^{[4,4]}(\mathbf k)-E_{\rm FB}^{(+)}
&=+4\zeta k^2,
&
E_{n,-}^{[4,4](+)}-E_{\rm FB}^{(+)}
&\simeq-\frac{3\zeta}{2\ell_B^2n}.
\end{align}
Thus, in both cases the anomalous Landau sequence lies on the side of the flat band energy opposite to the linear Landau sequence.

%\section{To be removed (but maybe keep for your thesis??}
%\redmark{{\large To be removed}: my current understanding of  the width of the $LL$  in momentum space in a TB calculation is the following: 
%\begin{itemize}
%\item in a TB calculation, the number of $k_x$ points determines the length of the ribbon in the {\em infinite} direction, $L_x$:  $N=\frac{2\pi/a}{\Delta k_x}$ or  $\Delta k_x = \frac{2\pi}{a N} =\frac{2\pi}{L_x}$, where $a$ is the length of the unit cell along the $x$ direction.
%\item The density of states $\rho $(number of states per unit area) of a given band is therefore: $\rho\equiv \frac{N}{L_x W}= \frac{\frac{L_x}{a}}{L_x W}=\frac{1}{aW}$
%\item The  degeneracy of a given Landau levels (see discussion above) is given by $N_{\rm state}= \frac{L_x L_y}{2\pi l_B^2} $
%\item Therefore, the number of states associated to a given Landau level, in a ribbon of width $W$, length $L_x$ should be $N_{\rm state}= \frac{L_x W}{2\pi l_B^2} $ and the density of states (number of states per unit area) of a given LL is $\rho_{LL}=\frac{1}{2\pi l_B^2}$

%\item Therefore, for a given band in a TB calculation, the fraction of the BZ occupied by the LL should be given by the ratio of densities $r=\frac{\rho}{\rho_{LL}}=\frac{1/aW}{1/2\pi l_ B^2}=2\pi\frac{l_B^2}{aW}$
%\end{itemize}
%In a given LL should be roughly be given by the ratio $W/l_B$.  The   Landau level states in a ribbon of width $W$ is }

\bibliography{LLlibfileup}

\end{document}